\documentclass[fleqn,usenatbib]{mnras}

\usepackage[T1]{fontenc}
\usepackage{ae,aecompl}

\usepackage{graphicx}	
\usepackage{amsmath}	
\usepackage{amssymb}	

\usepackage{orcidlink}

\newcommand{\GV}{green valley}
\newcommand{\RS}{red sequence}
\newcommand{\BC}{blue cloud}
\newcommand{\GZ}{Galaxy Zoo}

\usepackage{newtxtext,newtxmath}

\title[Green Valley Galaxy Morphology]{Galaxy And Mass Assembly: Galaxy Morphology in the Green Valley, Prominent rings and looser Spiral Arms}

\author[D. Smith et al.]{Dominic Smith$^{1}$\,\orcidlink{},     
Lutz Haberzettl$^{1}$\,\orcidlink{0000-0001-5620-3024},
L. E. Porter$^{1}$\,\orcidlink{0000-0002-7795-2267},
Ren Porter-Temple$^{1}$\,\orcidlink{0000-0001-7044-0632}, \and
Christopher P. A. Henry$^{1}$\,\orcidlink{0000-0002-2418-7151}, 
Benne Holwerda$^{1}$\,\orcidlink{0000-0002-4884-6756}\thanks{Corresponding author, E-mail: benne.holwerda@louisville.edu},
\'A. R. L\'opez-S\'anchez$^{2,3,4}$\,\orcidlink{0000-0001-8083-8046}, \and
Steven Phillipps$^{5}$\,\orcidlink{0000-0001-5991-3486},
Alister W. Graham$^{6}$\,\orcidlink{0000-0002-6496-9414}, 
Sarah Brough$^{7}$\,\orcidlink{0000-0002-9796-1363}, 
Kevin A. Pimbblet$^{8}$\,\orcidlink{0000-0002-3963-3919}, \and
Jochen Liske$^{10}$\,\orcidlink{0000-0001-7542-2927}, 
Lee S. Kelvin$^{11}$\,\orcidlink{0000-0001-9395-4759},
Clayton D. Robertson$^{1}$\,\orcidlink{0000-0002-5404-1372},
Wade Roemer$^{1}$\,, \and
Michael Walmsley$^{12}$\,\orcidlink{0000-0002-6408-4181}, 
David O'Ryan$^{13}$\,\orcidlink{} 
and Tobias G\'{e}ron$^{13}$\,\orcidlink{0000-0002-6851-9613}. 
 \\
$^{1}$ Department of Physics and Astronomy, University of Louisville\\
$^{2}$ Australian Astronomical Optics, Macquarie University, 105 Delhi Rd, North Ryde, NSW 2113, Australia\\
$^{3}$ Department of Physics and Astronomy, Macquarie University, NSW 2109, Australia\\
$^{4}$ ARC Centre of Excellence for All Sky Astrophysics in 3 Dimensions (ASTRO-3D)\\
$^{5}$ Astrophysics Group, School of Physics, University of Bristol, Tyndall Avenue, Bristol BS8 1TL, UK\\
$^{6}$ Centre for Astrophysics and Supercomputing, Swinburne University of Technology, Hawthorn, VIC 3122, Australia\\
$^{7}$ School of Physics, University of New South Wales, NSW 2052, Australia\\
$^{8}$ E.A.Milne Centre for Astrophysics, University of Hull, Cottingham Road, Kingston-upon-Hull, HU6 7RX, UK\\
$^{10}$ Hamburger Sternwarte, Universit\"at Hamburg, Gojenbergsweg 112, 21029 Hamburg, Germany\\
$^{11}$ Department of Astrophysical Sciences, Princeton University, 4 Ivy Lane, Princeton, NJ 08544, USA\\
$^{12}$ Jodrell Bank Centre for Astrophysics, Department of Physics \& Astronomy, University of Manchester, Oxford Road, Manchester M13 9PL, UK\\ 
$^{13}$ Observational Astrophysics Group, Lancaster University, LA1 4YW, UK\\
$^{14}$ Department of Physics, University of Oxford, Denys Wilkinson Building, Keble Road, Oxford OX1 3RH, UK
}

\date{Accepted XXX. Received YYY; in original form ZZZ}

\pubyear{2015}

\begin{document}
\label{firstpage}
\pagerange{\pageref{firstpage}--\pageref{lastpage}}
\maketitle

\begin{abstract}
Galaxies broadly fall into two categories: star-forming (blue) galaxies and quiescent (red) galaxies. In between, one finds the less populated ``green valley". Some of these galaxies are suspected to be in the process of ceasing their star-formation through a gradual exhaustion of gas supply or already dead and are experiencing a rejuvenation of star-formation through fuel injection. We use the Galaxy And Mass Assembly database and the Galaxy Zoo citizen science morphological estimates to compare the morphology of galaxies in the green valley against those in the red sequence and blue cloud. 
Our goal is to examine the structural differences within galaxies that fall in the green valley, and what brings them there. Previous results found disc features such as rings and lenses are more prominently represented in the green valley population. We revisit this with a similar sized data set of galaxies with morphology labels provided by the Galaxy Zoo for the GAMA fields based on new KiDS images. Our aim is to compare qualitatively the results from expert classification to that of citizen science. 

We observe that ring structures are indeed found more commonly in green valley galaxies compared to their red and blue counterparts. We suggest that ring structures are a consequence of disc galaxies in the green valley actively exhibiting characteristics of fading discs and evolving disc morphology of galaxies. We note that the progression from blue to red correlates with loosening spiral arm structure. 

\end{abstract}




\section{Introduction} \label{sec:intro}

Previous large-scale surveys of galaxies have revealed a bi-modality in the colour-magnitude diagram of \mbox{galaxies} with two distinct populations: one with blue optical colours and another with red optical colours \citep{Strateva01, Baldry04, Bell04, Martin07, Faber07,Baldry06, Willmer06, Ball08, Brammer09,Mendez11,Taylor15,Corcho-Caballero20,Corcho-Caballero21}. These populations were dubbed the ``blue cloud'' (BC), or ``star-forming galaxy sequence'', and the ``red sequence'' (RS) respectively \citep{Driver06, Faber07, Salim15}. The blue cloud and red sequence are best separated at higher stellar mass and mix at lower stellar masses \citep[cf.][]{Taylor15}.


The Galaxy Zoo (GZ) project \citep{Lintott08}, which produced morphological classifications for a million galaxies, helped to confirm that this bi-modality is not entirely morphology driven \citep[][Figure \ref{fig:gz:questions}]{Salim07, Schawinski07, Bamford09, Skibba09}. It suggested larger fractions of spiral galaxies in the \RS{}\footnote{The red sequence was originally known as the color-magnitude relation for early-type galaxies, see the review in \cite{Graham13}.} \citep{Masters10} and elliptical galaxies in the \BC{} \citep{Schawinski09} than had previously been detected.

The sparsely populated colour-mass space between these two populations, the so-called ``Green Valley'' (\GV) (Figure \ref{fig:colour-mass}), provides clues to the nature and duration of a galaxies’ transitions from \BC{} to \RS{}. This transition must occur on rapid timescales, otherwise there would be an accumulation of galaxies residing in the \GV{}, rather than an accumulation in the \RS{} as is observed \citep{Arnouts07, Martin07, Smethurst15,Smethurst17,Nogueira-Cavalcante17,Bremer18,Phillipps19,Barone22}. Alternatively, gas infall onto \RS{} galaxies may rejuvenate them into the \GV{} \citep[e.g.][]{Graham17}. 
\GV{} galaxies have therefore long been thought of as the ``crossroads'' of galaxy evolution, a transitional population between the two main galactic stages of the star-forming \BC{} and the ``red \& dead'' sequence \citep{Bell04, Martin07, Faber07, Mendez11, Schawinski14, Pan15,Graham19b}, however, it is possible that these are also \RS{} galaxies that have been rejuvenated \citep{Graham15,Graham17}.

The intermediate colours of these \GV{} galaxies have been interpreted as evidence for recent quenching (suppression) of star formation \citep[][]{Salim07, Salim15, Smethurst15,Phillipps19}. Star-forming galaxies are observed to lie on a well defined stellar mass-SFR relation \citep{Martin05}, however, quenching a galaxy causes it to depart from this relation \citep{Noeske07, Peng10}. The main mechanism for galaxy quenching is thought to be a lack of fuel for star formation. Fading of the star-forming disc, the primary site of star formation, drives the apparent morphological transition of galaxies from spiral to lenticular or elliptical in the galaxies that are quenching and lie in the \GV{}  \citep{Coenda18a,Bluck20,Fraser-McKelvie19b,Fraser-McKelvie20a,Fraser-McKelvie20b}.

\cite{Kelvin18} examined 472 galaxies in the Galaxy And Mass Assembly (GAMA) survey with SDSS imaging visually for signs of disc substructures (e.g., rings, bars, and lenses) with a team of expert classifiers. They found evidence that rings and lenses are more common in the \GV{} then in \RS{} and \BC{} galaxies. Our goal here is to re-examine this result using the GZ morphological estimates using the higher resolution and deeper Kilo-Degree Survey (KiDS) images \citep{Kuijken19} of the galaxies in the GAMA survey's equatorial fields \citep{Driver11}. Our sample is of similar size as that of \cite{Kelvin18} (396 vs 472) and there is likely overlap. There are two critical differences: the method of classification and the quality of the data. Our classifications are based on much improved data and arrived at with citizen science voting rather than a small expert panel. Our aim is to examine if the different data and classification schemes arrive at qualitatively the same conclusions for the morphology of the \GV{}.

Our paper is organized as follows: Section \ref{sec:data} describes the GAMA and GZ data we use here, Section \ref{sec:results} presents our results and we discuss these in Section \ref{sec:discussion}. Section \ref{sec:conclusions} lists our conclusions.

\section{Data} \label{sec:data}

In this paper, we use GAMA and Galaxy Zoo voting data based of the KiDS imaging. 

\subsection{GAMA} \label{sec:gama}

The GAMA survey is a  spectroscopic survey comprised of three equatorial fields and two Southern fields. Its multi-wavelength photometry ranges from ultraviolet to sub-mm wavelengths \citep{Driver09,Hopkins13g,Liske15}. Redshifts (z) are reliably found to z$\sim0.8$ and the survey is complete to $\sim$98\% for an apparent magnitude in Sloan Digital Sky Survey $r$-filter (SDSS-$r$) of 19.8 mag in the equatorial fields. Here, we consider only these (G09, G12 and G15), as they overlap with the KiDS \citep[KiDS][]{de-Jong13,de-Jong15,de-Jong17,Kuijken19}, on which the GZ information is based (see next section). 

We use GAMA optical photometry from SDSS Stripe-82 photometry  \cite{Taylor11}, which is corrected for redshift (K-correction) and internal dust extinction.
The final photometry we used is the LAMBDAR photometry (v01) presented in \cite{Wright17}.

Separately, we use stellar masses derived from the spectral energy distribution (SED) model fit using {\sc magphys} tool \citep{da-Cunha08}, presented in \cite{Driver16,Wright17} (v06 in DR3). 

\begin{figure*}
    \centering
    \includegraphics[width=\textwidth]{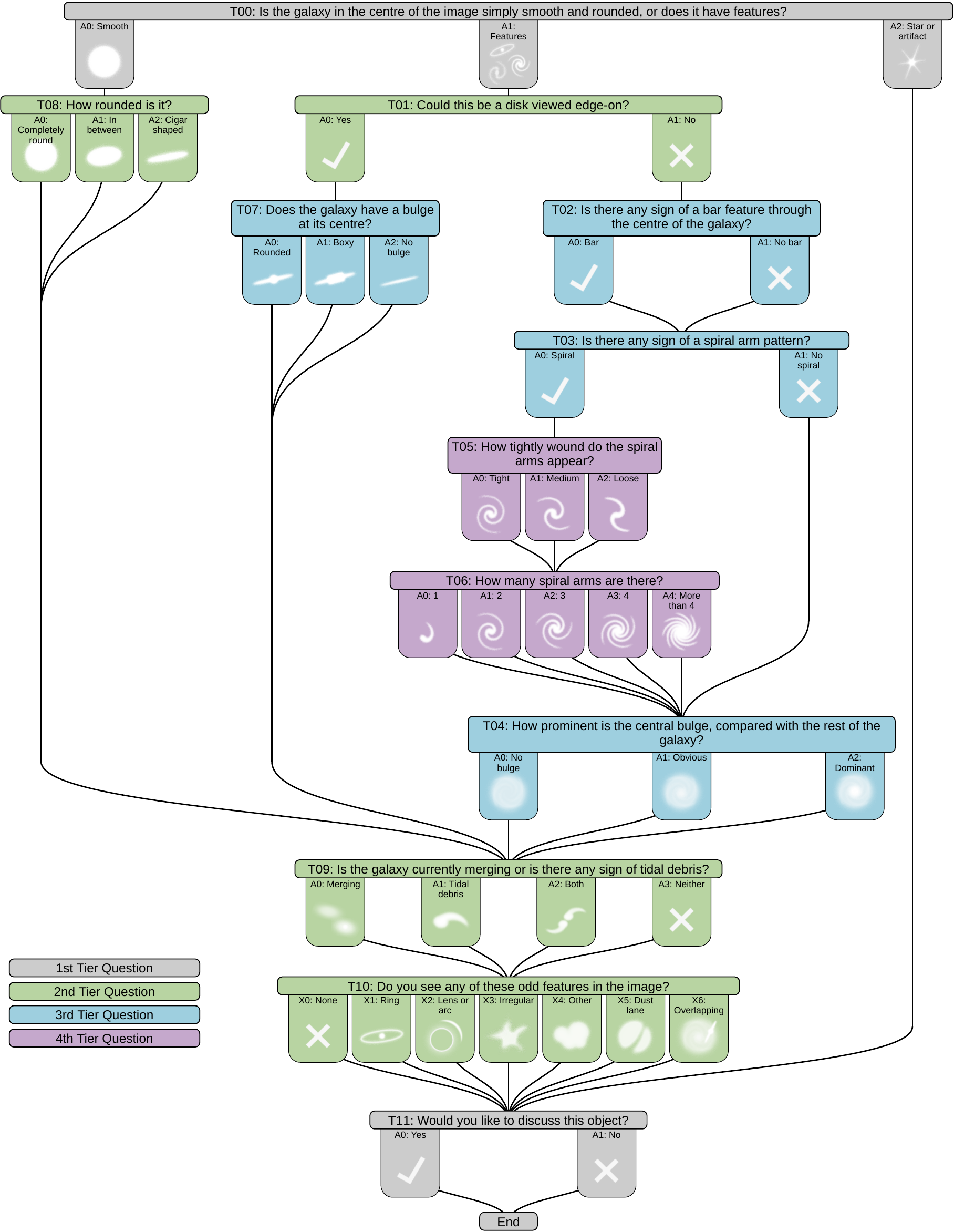}
    \caption{The flow diagram of the GZ4 (fourth generation) question tree. We refer to the text for details on the questions in the GAMA sample.}
    \label{fig:gz:questions}
\end{figure*}

\subsection{KiDS GZ} \label{sec:gz}

The GZ citizen science project analyzed KiDS \citep{de-Jong13,de-Jong15,de-Jong17,Kuijken19} images. Citizen scientists answered a series of questions based on KiDS g-band and r-band imaging. A synthetic green channel was constructed as the arithmetic mean of the other two to allow for the construction of three-colour RGB images. Because morphological detail is lost with distance, a limit of $z\leq 0.075$ is enforced to ensure reliable morphological estimates of kpc-scale structures (e.g. spiral arms, bars). Initially we ran this data with a $z\leq0.15$, however, we realized limitations in the Galaxy Zoo data that limited us to $z\leq0.08$ making resolution for many morphological structures difficult. Therefore, we decided on our sample to $z\leq 0.075$ to limit the bias due to distance effects in the Galaxy Zoo voting.

The GZ question tree is presented in \cite{Holwerda19} and in Figure \ref{fig:gz:questions}. The full Galaxy Zoo classification is described in Kelvin et al. (\textit{in preparation}). We focus on the questions that are asked in the GZ question tree (Figure \ref{fig:gz:questions}) regarding disc galaxy morphology. In section 3, we begin each subsection with the question code and associated question of the morphological features as asked in the GZ questionnaire. These question codes are T00, T01, T02, T03, T04, T05, T06, T09 and T10. 
We refer the reader to Kelvin et al. (\textit{in preparation}) for specific details of the GZ analysis. We used an internal GAMA/KiDS catalogue for the subsequent analysis. 

\begin{figure*}
    \centering
    \includegraphics[width=0.49\textwidth]{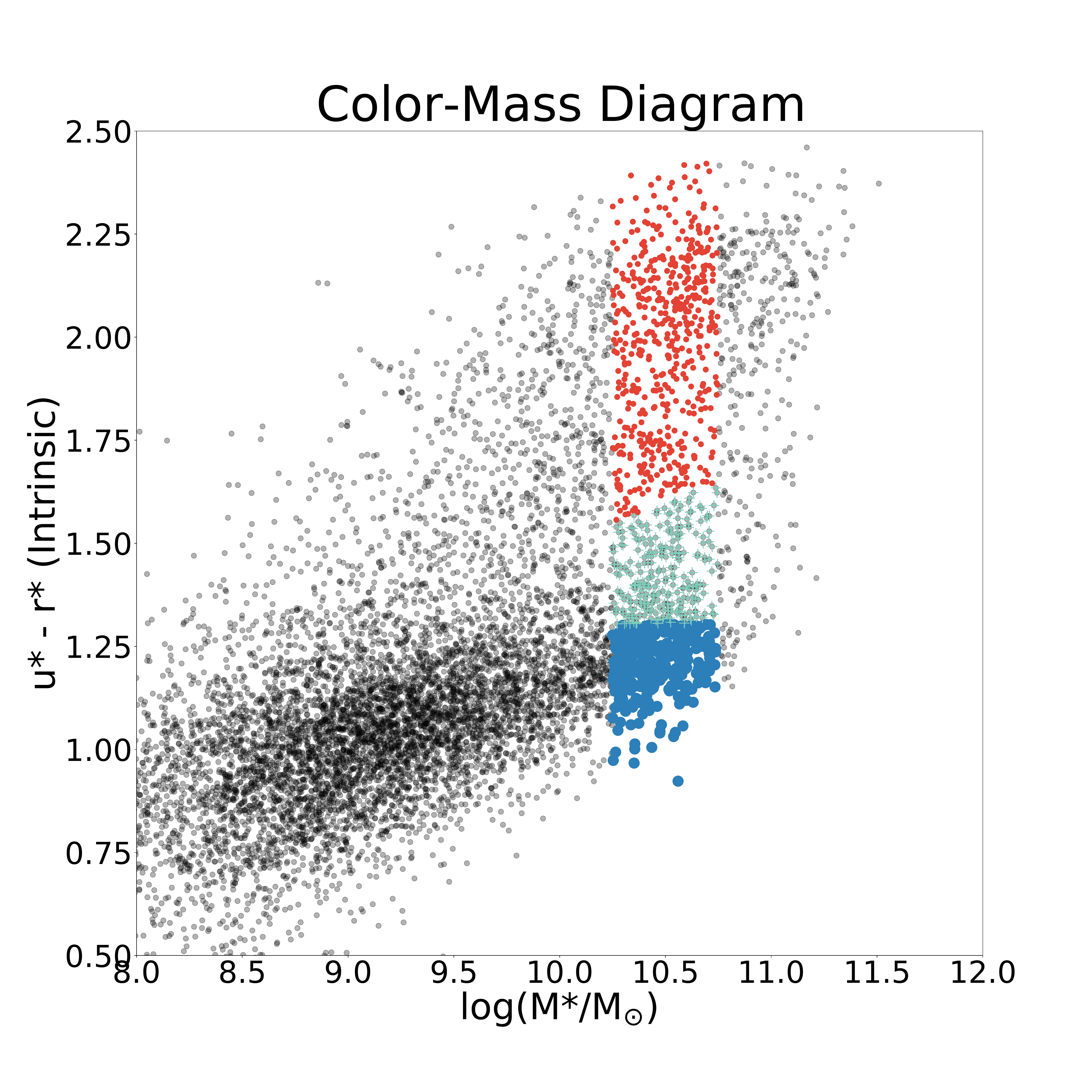}
    \includegraphics[width=0.49\textwidth]{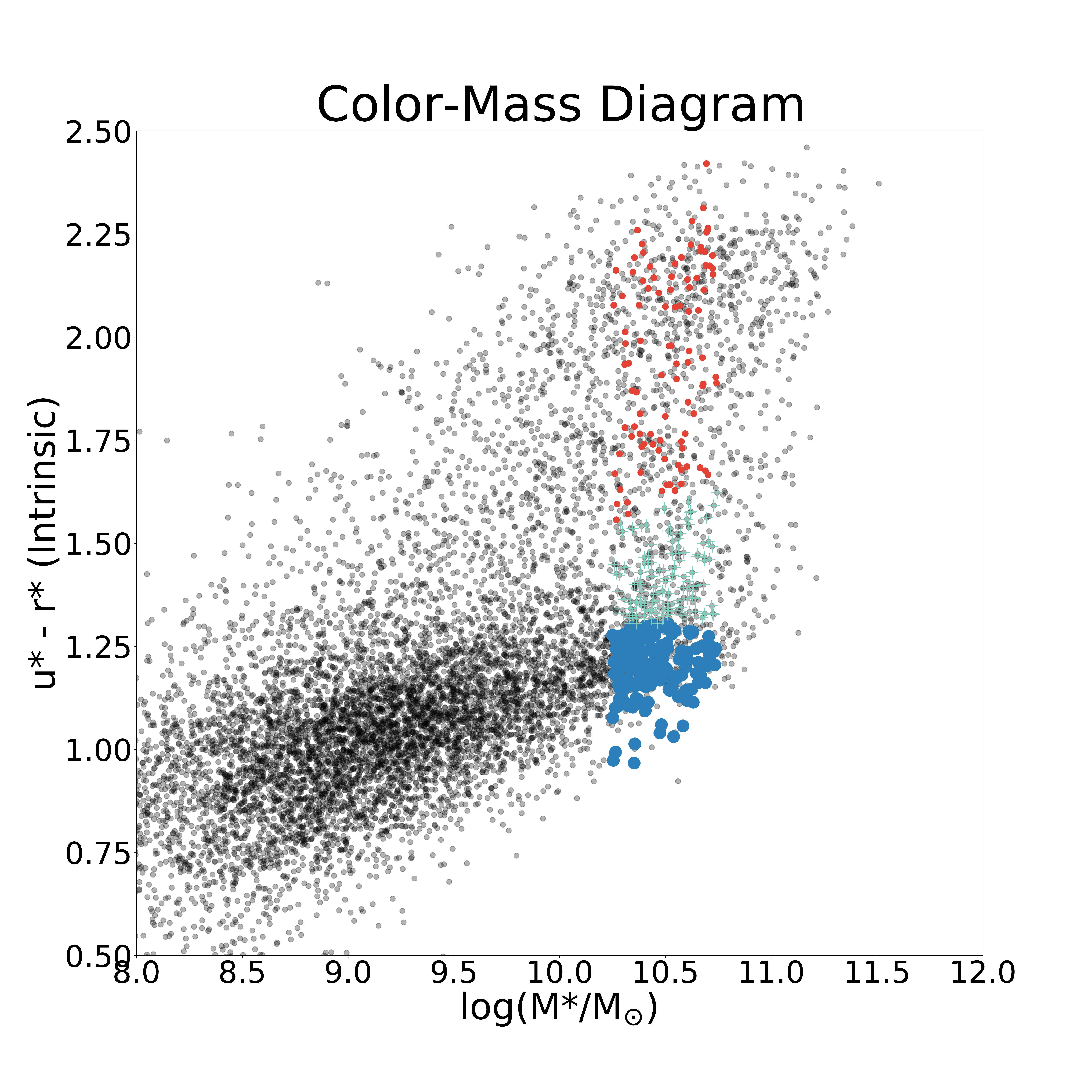}
    \caption{
    These plots represent all the GAMA galaxies in our mass range (10.25 $<$ log($M_*/M_{\odot}$) $<$ 10.75) colour -coded for their classification. We use the limits from \protect\cite{Bremer18} to select red, green and blue galaxies. 
    The left panel represents all the respective galaxies in their mass ranges before further data selection. The right panel shows all galaxies with any votes between .1 and 1 for not being seen edge on and votes between .3 and 1 for spiral features. The right image represents our current data selection. }
    \label{fig:colour-mass}
\end{figure*}

\begin{figure}
    \centering
    \includegraphics[width=0.5\textwidth]{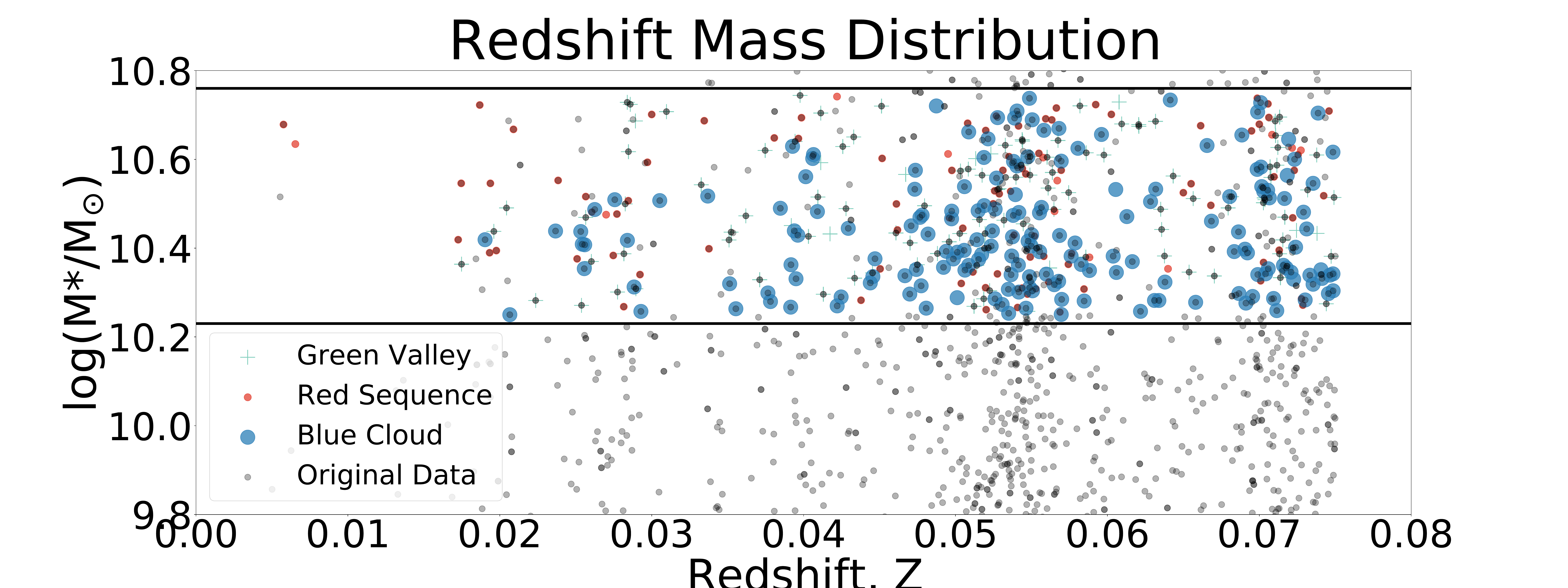}
    \caption{A scatter plot of the blue, green, and red galaxies in our sample extracted from GAMA to compare to GZ voting. The coloured galaxies represent the 396 galaxies that composed our selected sample represented by the right graph of Figure \protect\ref{fig:colour-mass}. The GAMA sample is taken for $10.25 < log(M_*/M_{\odot}) < 10.75$ at z $\leq$ 0.075, which is the redshift limit of the GZ selection from KiDS. The colour criteria are from \protect\cite{Bremer18}.}
    \label{fig:z-m}
\end{figure}

\begin{figure}
    \centering
    \includegraphics[width=0.5\textwidth]{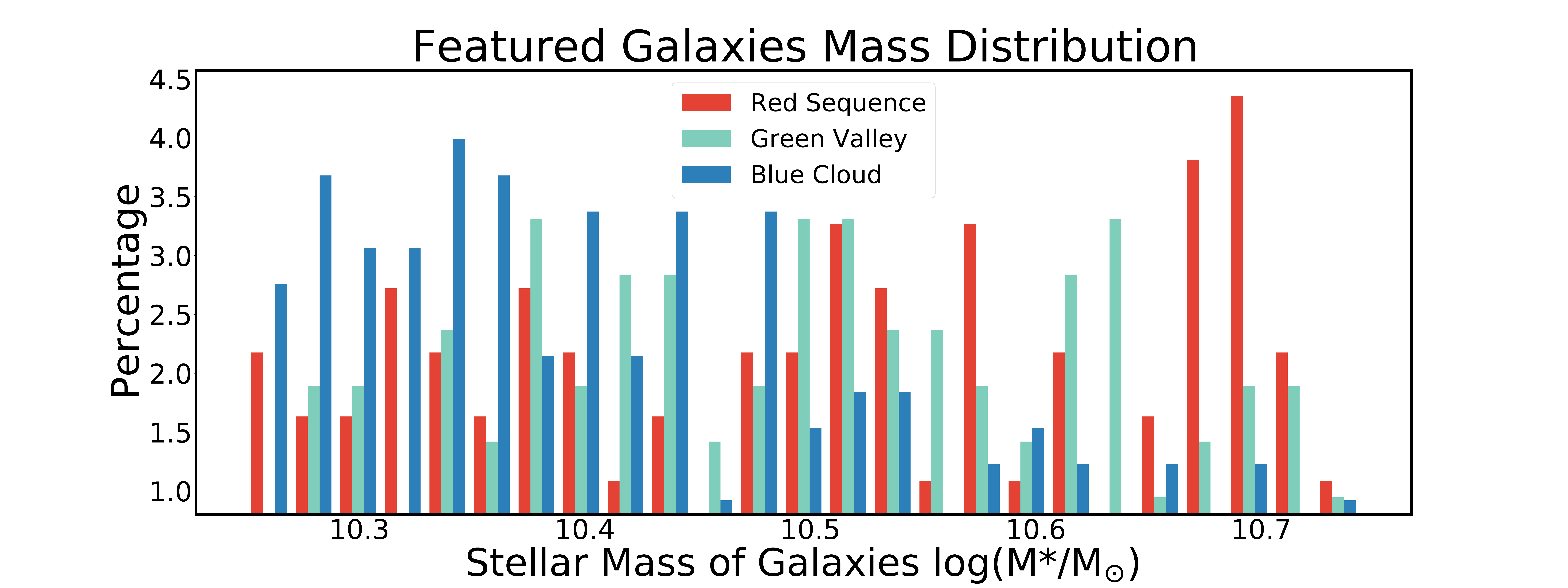}
    \includegraphics[width=0.5\textwidth]{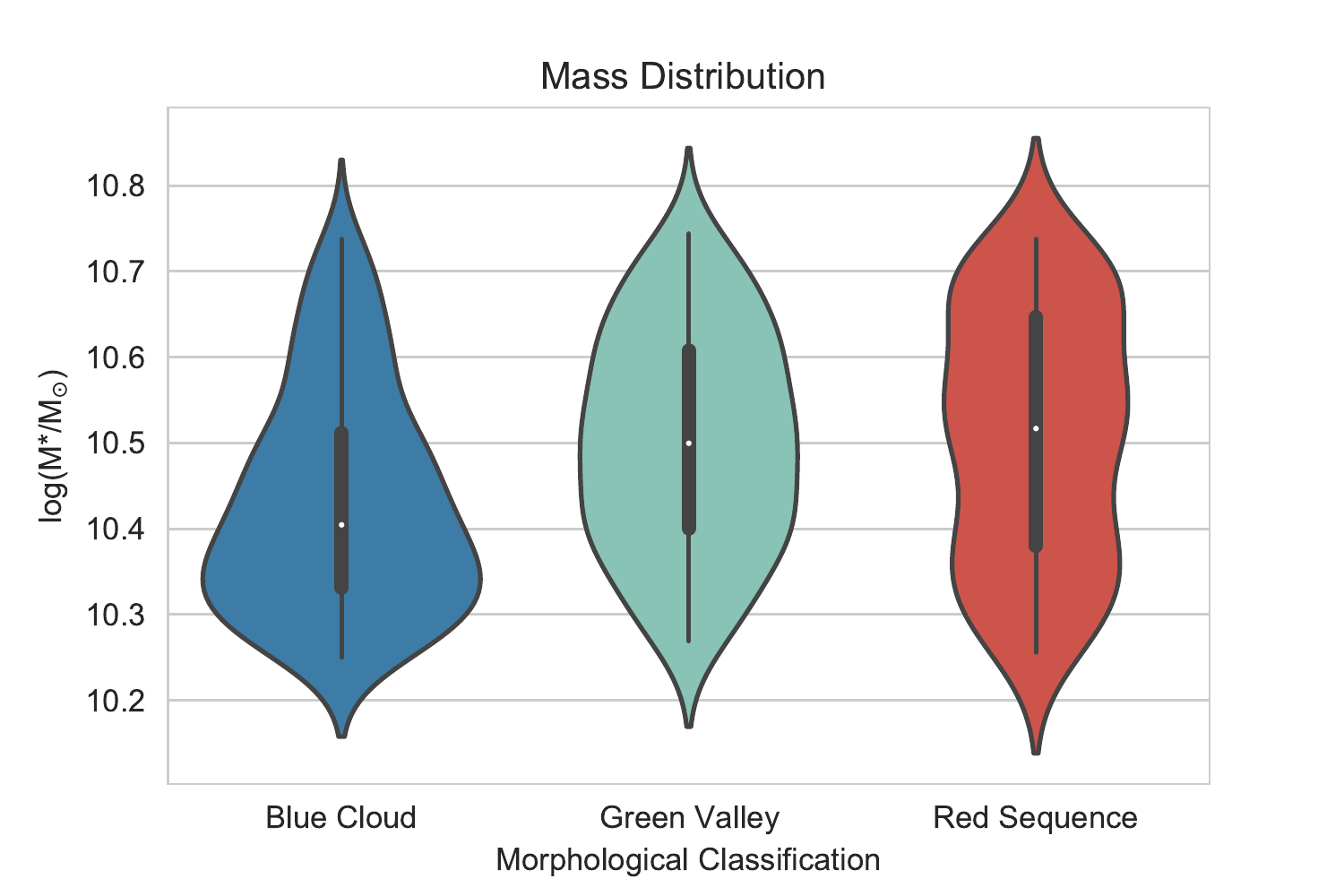}
    \caption{A histogram of the blue, green, and red galaxies in our selection sample ( right graph of Figure \ref{fig:colour-mass}) extracted from GAMA to compare with GZ voting. The GAMA sample is taken for (10.25 $<$ log($M_*/M_{\odot}$) $<$ 10.75) and z $\leq$ 0.15, the limit of the GZ selection from KiDS. The histogram illuminates the overlap of data presented in Figure \ref{fig:z-m}. The violin plot below shows the same information as in the histogram above it. }
    \label{fig:z-m_hist}
\end{figure}

\subsection{Sample selection}

We organized our GZ data by stellar mass, log($M_*)/M_{\odot}$, vs. intrinsic stellar population colour plane ($u$* - $r$*) \citep[the population selections are from][]{Bremer18}. Figure \ref{fig:colour-mass} shows our selection of \BC{}, \GV{}, and \RS{} galaxies, based on their stellar mass and rest-frame colour. 
A second requirement is that these galaxies are disc-dominated following the T00 question (Figure \ref{fig:gz:questions}) ``Is the galaxy in the centre of the image simply smooth and rounded or does it have features?'' with any fraction of the votes in favor of any features ($f_{features} > 0$). This is to remove any galaxies without features to further examine. 

For the first graph in Figure \ref{fig:colour-mass} we limited the mass to 10.25 $<$ log($M_*/M_{\odot}$) $<$ 10.75 and a redshift of $z \leq 0.075$. This mass range was selected to ensure a complete sample and for a more direct comparison with \cite{Kelvin18}, who select galaxies in the same mass range. 
This mass range covers the tip of the blue cloud at low z, this is necessary because for galaxies to continue evolving beyond this point, they must transition across the green valley \citep[][]{Bremer18}. The redshift limit was imposed to ensure that distance effects on the GZ voting are minimal \citep[see the discussion in][on distance effects]{Willett13}. Beyond z=0.075, the resolution of KiDS images is not sufficient to discriminate features of a kiloparsec in scale, such as the width of spiral arms and rings. The voting can be corrected using debiasing but the GAMA dataset may be too small for that (but the Galaxy Zoo v4 will be). 

The right graph of Figure \ref{fig:colour-mass} represents the galaxies in each faction that are voted in the GZ to have any votes for ``Could this be a disc viewed edge on'' as no and ``Is there any sign of a spiral arm pattern?" that registered 30\% or higher. (T01 and T03 in the Figure \ref{fig:gz:questions} respectively). To do this, we set the voting for T01 results to fraction limits between 0.0001 and 0.9999 to prevent the possible error of galaxies with single votes throwing off the results, and to maximize those votes for features. We continued to do the same with T02, however this time we set the limits between .3 and 1. This allowed us to choose galaxies that had votes for bulges, spiral features and all other morphological features down this branch of the questionnaire. Lastly for this step we limited our selection for spiral features 
in the same way.  We were left with a data set of 176 for \BC{}, 118 for \GV{}, and 102 for \RS{} for the total of 396 in the mass range considered.

Figure \ref{fig:z-m} shows the distribution of featured galaxies in redshift (z) and stellar mass ($M_*$). Our selection is made to the specifications of 10.25 $<$ log($M_*/M_{\odot}$) $<$ 10.75 and z $\leq$ 0.075, the redshift limit of GZ pre-selection for classification. This gives us a good representative volume to compare galaxy morphology. We find that \RS{} galaxies are at slightly higher masses than the \BC{}, with the \GV{} galaxies spanning an intermediate mass range. The galaxy spread is better represented by mass in Figure \ref{fig:z-m_hist}.

\section{Results} \label{sec:results}

We compared the normalized fractions of  galaxies between the \BC{}, \GV{}, and \RS{} galaxies to generate the voting histograms and violin plots showing sample fractions in the following subsections.

As shown in Figure \ref{fig:gz:questions}, each tier signified with a code, T\#\#. This code represents question asked in the GZ about a morphological trait with the number denoting the tier of each question. We organized each subsection by these question codes with their represented data in a histogram. 

The tool used to compare these data is the Kolmogorov-Smirnov (K-S) two-sample similarity test and the associated p-value  calculated for our selection samples. The K-S value provides the maximum difference between any 2 cumulative distributions we consider. The p-value is the probability of  the random occurrence of the presented null hypothesis (the distributions are the same). 

{We plot the voting in the Galaxy Zoo in two ways in each of the following Figures. In the top panel, we plot the cumulative histogram of the voting in Galaxy Zoo: on the x-axis is the fraction of voting in favor of the question under consideration and on the y-axis the fraction of the sample is shown. A plot that rises early has a larger fraction of the sample with a low fraction of the votes in favor of this feature being present. In this case, the feature is relatively rare. If the plot rises on the right of the x-axis, a large fraction of the sample has a high fraction of votes (or greater consensus) that this feature is present. }

{In the lower panel we present the same fraction of the voting in a more traditional histogram, rendered as a violin plot (a mirrored histogram with a kernel density applied to render it into a smooth graph). Because the distribution is slightly smoothed, the range of values in the y-axes in the Figures goes from -0.2 to 1.2 to accommodate the tails resulting from the kernel smoothing. The range and standard deviations of the distribution are also shown as thin and thick horizontal lines.}

{By combining both graphical visualizations in each plot, we hope to show both when voting behaviour between populations is similar or dissimilar in the cumulative distribution, reflected in the K-S metric, and how the voting behavior looks in each population in a more intuitive histogram rendering. }

\begin{figure}
    \centering
    \includegraphics[width=0.5\textwidth]{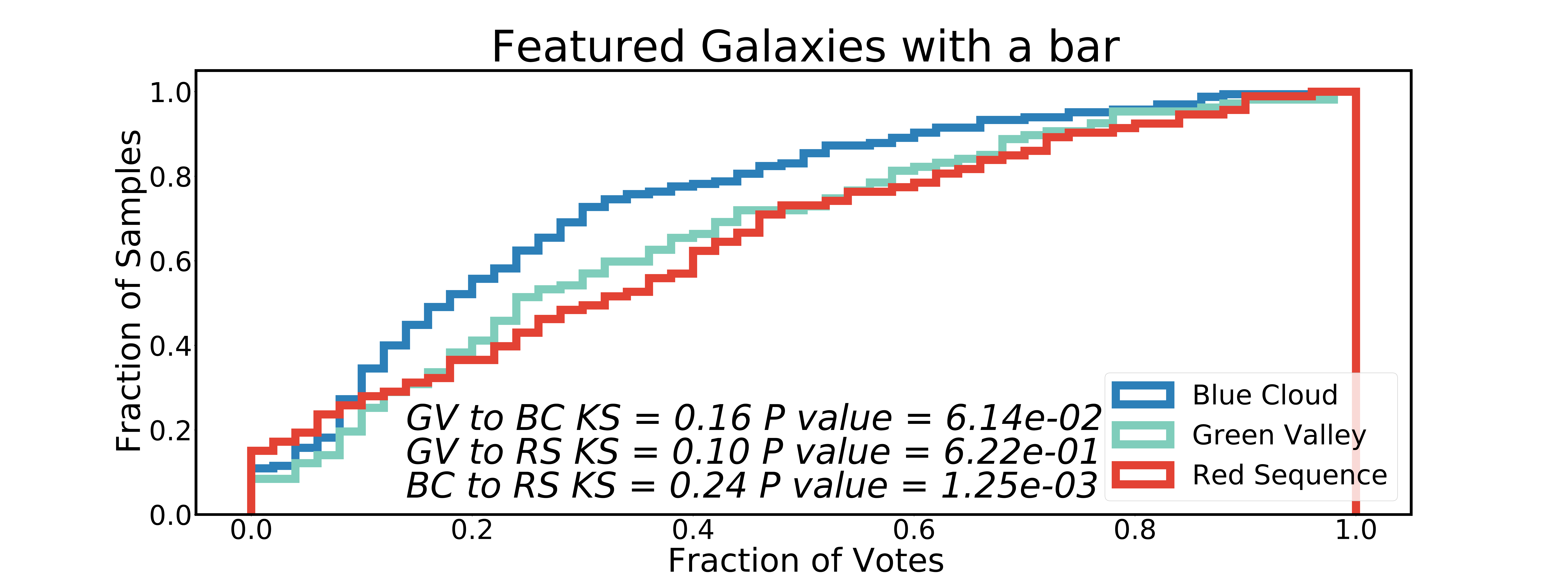}
    \includegraphics[width=0.5\textwidth]{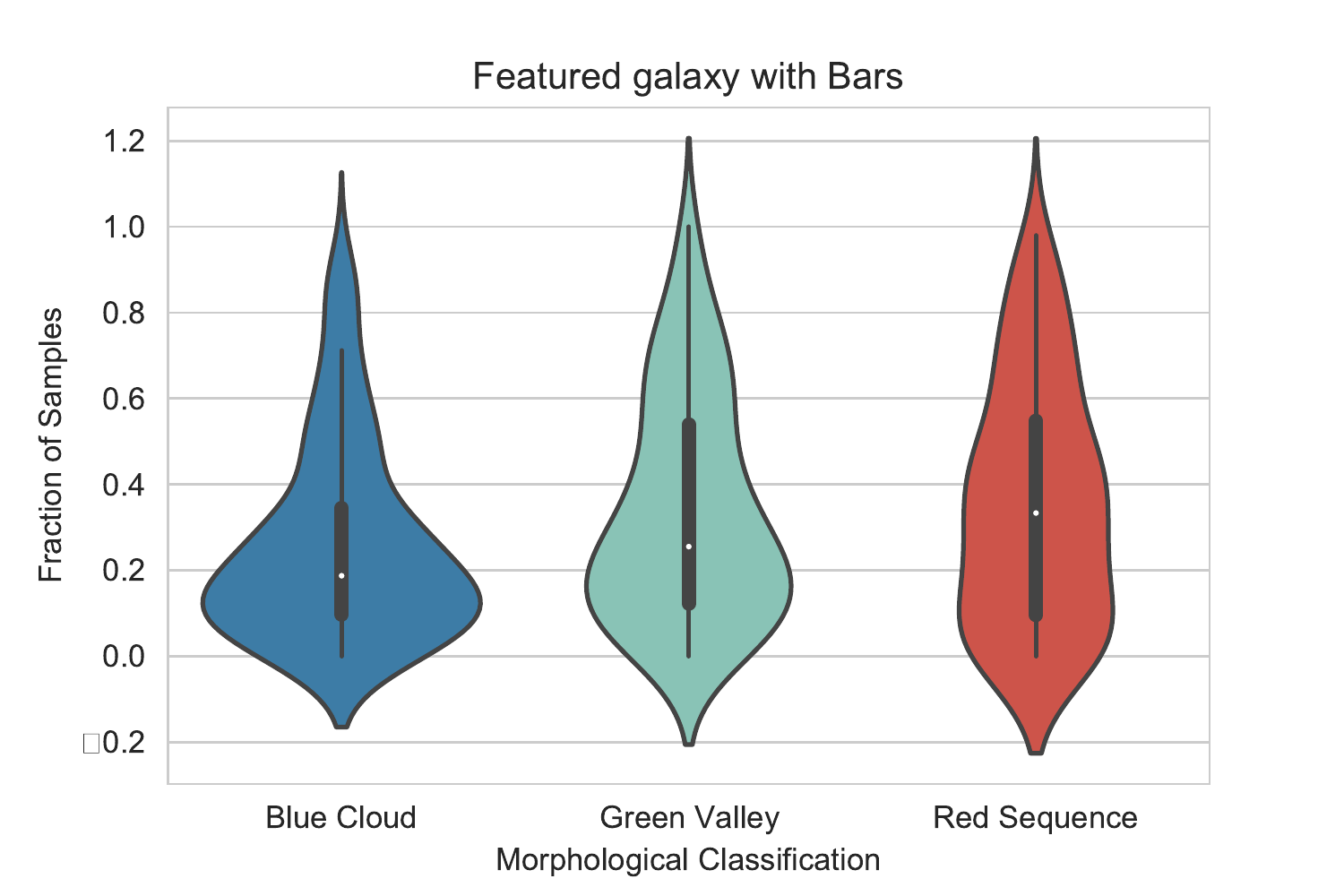}
   \caption{A histogram of the fraction of votes in favor of classifying galaxies as featured with a bar (T02 in the GZ questionnaire). The difference between the 3 groups is nearly equally as distinguishable with K-S values of 0.16 for \GV{} and \BC{}, 0.10 for \GV{} and \RS{}, and 0.24 for \BC{} and \RS{}. The significance from p-values is $6.14\times10^{-02}$, $6.22\times10^{-01}$, and $1.25\times10^{-03}$ respectively. This confirms that the 3 groups are statistically variant. The violin plots represent the same data}
    \label{fig:hist:bar}
\end{figure}

\subsection{Bars}

Figure \ref{fig:gz:questions}, T02: \textit{``Is there a sign of a bar feature through the centre of the galaxy?"}

Stellar bars are a prime suspect for a morphological feature that aids in quenching, especially quenching from the inside-out \citep[see][for a review]{Masters21}.

GZ voting show that \RS{} and \GV{} have a lower fraction of galaxies having bar shaped structures than \BC{}. In our statistical analysis, we have chosen to only include galaxies that were voted as bar galaxies 50 \% ($f_{bar} > 0.5$) or more of the time. From our galaxy sample for each faction, 12.1\% of \BC{}, 20.1\% of \GV{}, and 19.3\% of \RS{} galaxies met this requirement. These percentages are to be expected if the \GV{} is a transition zone and bars are in fact a predecessor to quenching and long-lived enough to do so. Figure \ref{fig:hist:bar} shows the similar voting behaviour in Galaxy Zoo in the \RS{} and \GV{}. The notably different behaviour in the \BC{} shows lower confidence in more of the \BC{} sample of galaxies.

\subsection{Featured discs}

\begin{figure}
    \centering
    \includegraphics[width=0.5\textwidth]{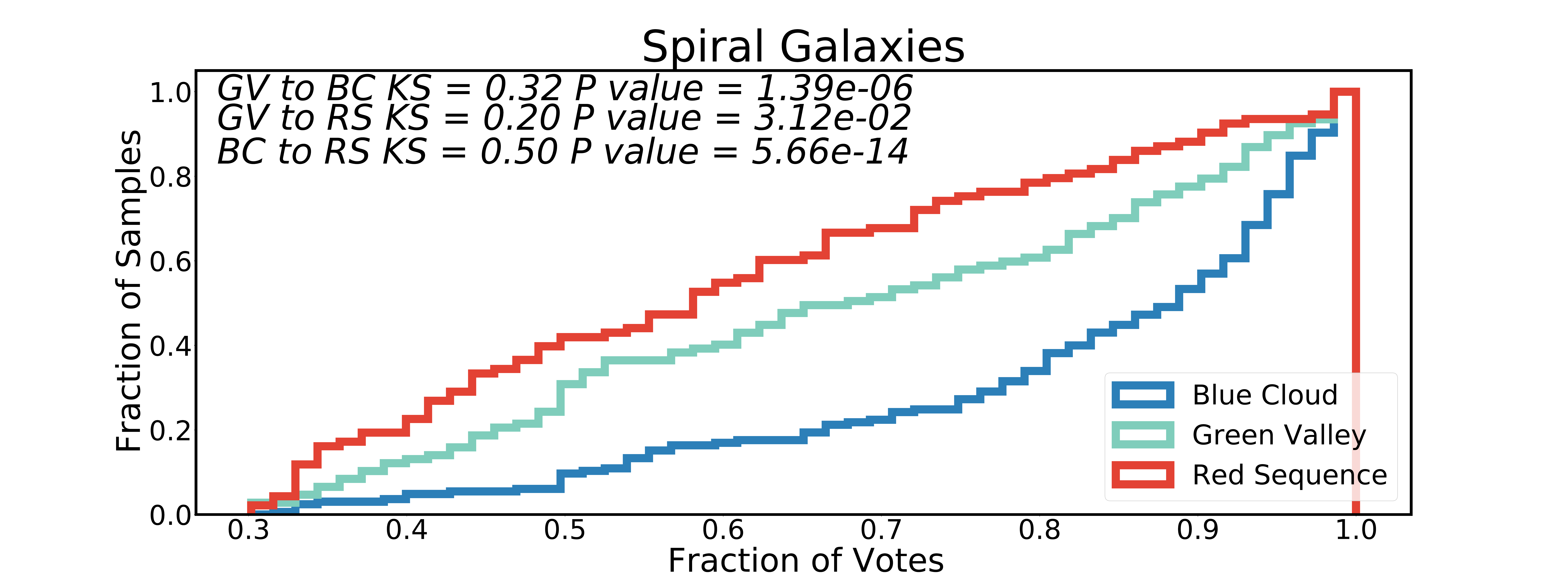}
    \includegraphics[width=0.5\textwidth]{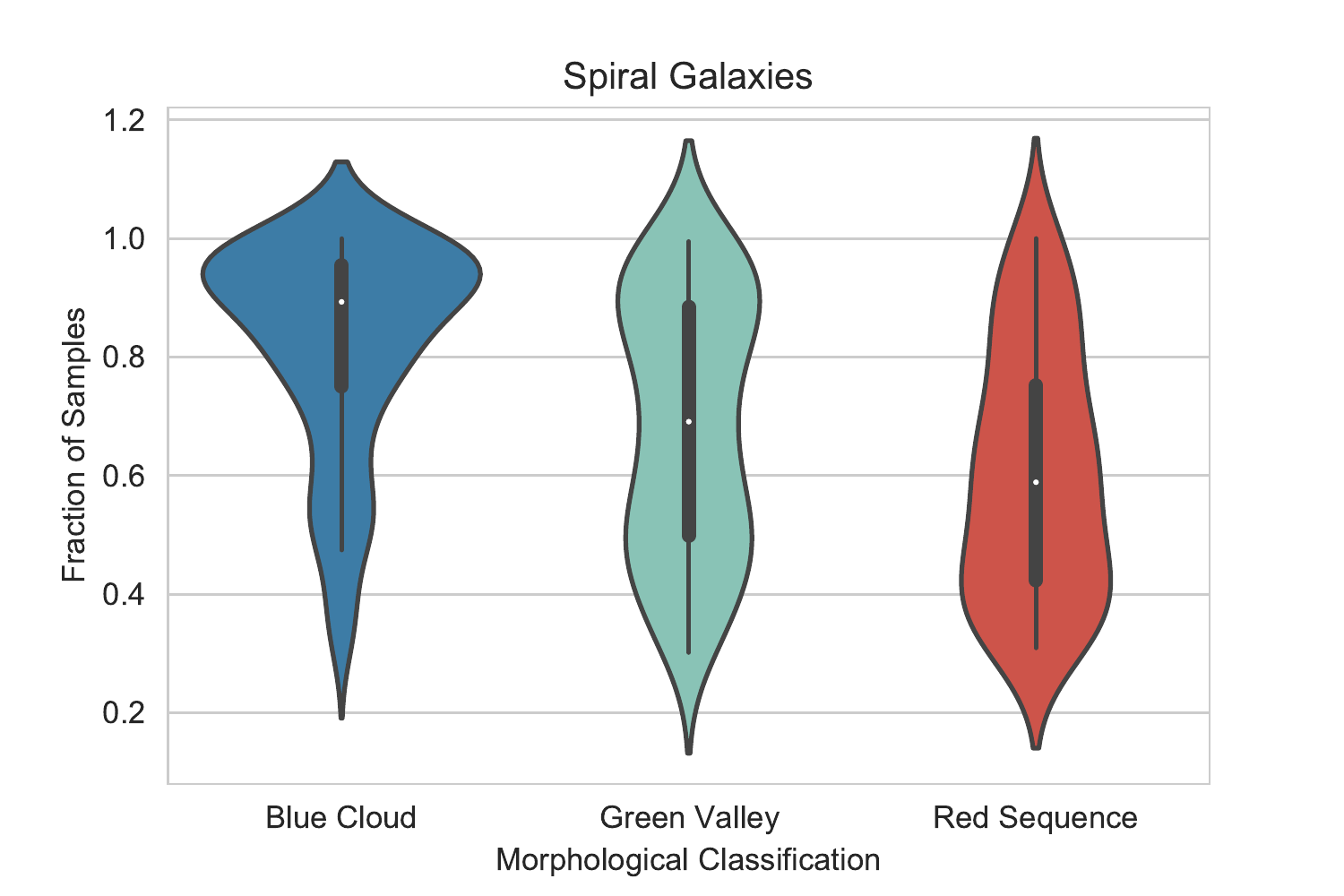}
    \caption{A histogram of the fraction of votes in favor of classifying galaxies as featured (T00 in the GZ questionnaire). According to the K-S test, the difference between the \GV{} and \BC{} is very distinguishable with a value of 0.37. The difference between \GV{} and \RS{} is distinguishable as well, with a KS value of 0.20. The difference between \BC{} and \RS{} results from a KS value of 0.50. These differences are further confirmed due to very small p-values. They are $1.39\times10^{-06}$, $3.12\times10^{-02}$, $5.66\times10^{-14}$ respectively. The violin plots represent the same data as the cumulative histogram at the top.}
    \label{fig:hist:spiral}
\end{figure}

Figure \ref{fig:gz:questions}, T03: \textit{``Is there any sign of a spiral arm pattern?"}

Using the voting data from this question showed the galaxies with spiral features in the \RS{}, \BC{}, and \GV{}. 

The outcome, as shown in Figure \ref{fig:hist:spiral}, shows that most disc galaxies in the \BC{} are featured galaxies. There are fewer featured galaxies among the disc galaxies of the \RS{}. The number of featured galaxies in the \GV{} falls somewhere in between those present in the \BC{} and \RS{}.   
The fractions of galaxies with discs is possibly an under-estimate as lenticular galaxies are often missed in visual inspections \citep[][]{Graham19b}.

\begin{figure}
    \centering
    \includegraphics[width=0.5\textwidth]{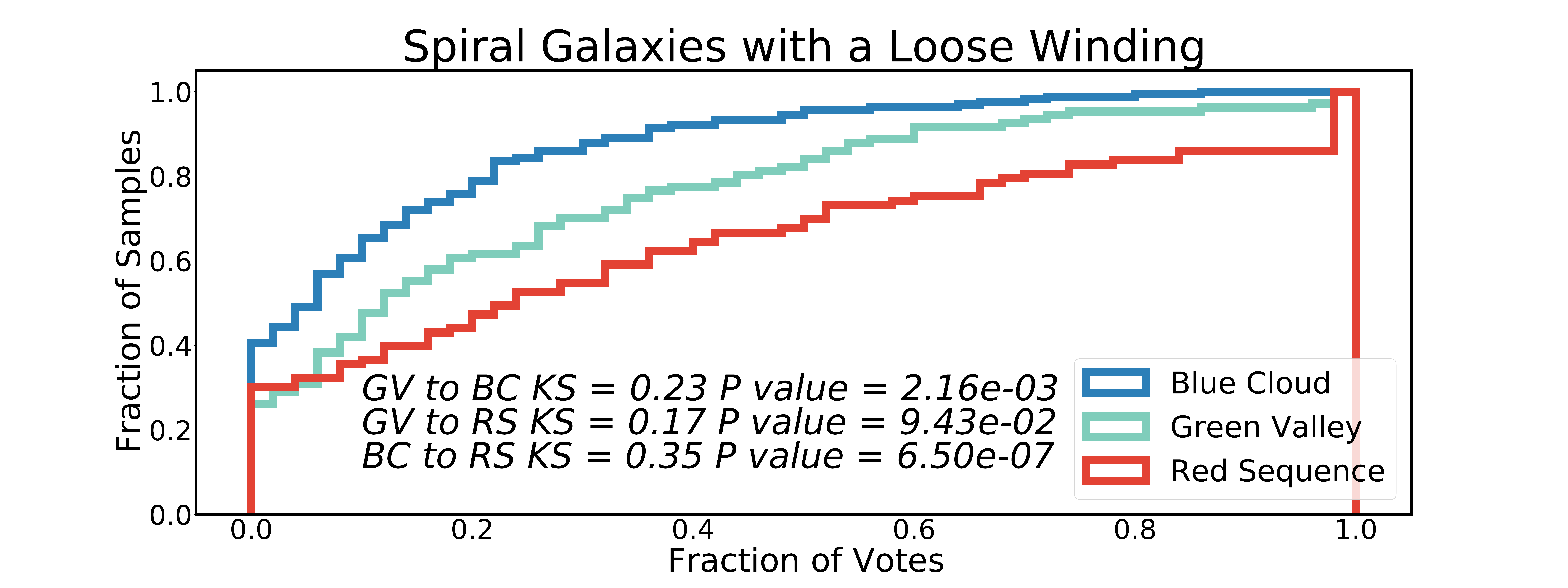}
    \includegraphics[width=0.5\textwidth]{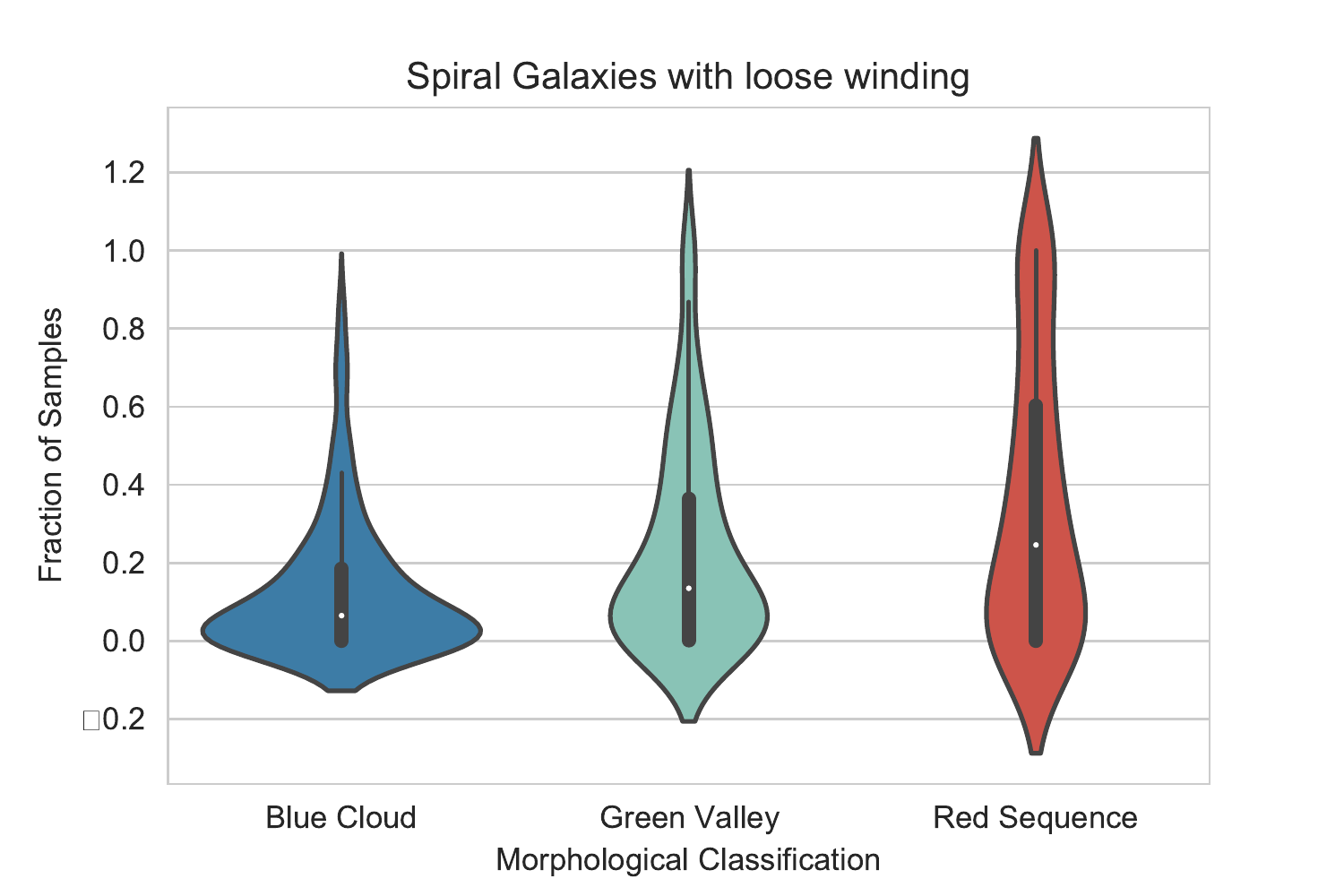}
    \caption{Histogram of the fraction of votes in favor of the spiral galaxies having a Loose winding  (T05 in the GZ questionnaire). The difference between \GV{} and \BC{} galaxies is distinguishable with a KS of 0.23, the KS of \GV{} and \RS{} is 0.17, and the KS of the \BC{} and \RS{} is 0.35. \GV{} to \BC{} is at a p value $2.16\times10^{-03}$,  \GV{} and \RS{} is $9.43\times10^{-02}$, and the KS of \RS{} and \BC{} is $6.5\times10^{-07}$.The violin plots in the lower panel represent the same data as the cumulative histogram at the top.}
    \label{fig:hist:loose}
\end{figure}

\begin{figure}
    \centering
    \includegraphics[width=0.5\textwidth]{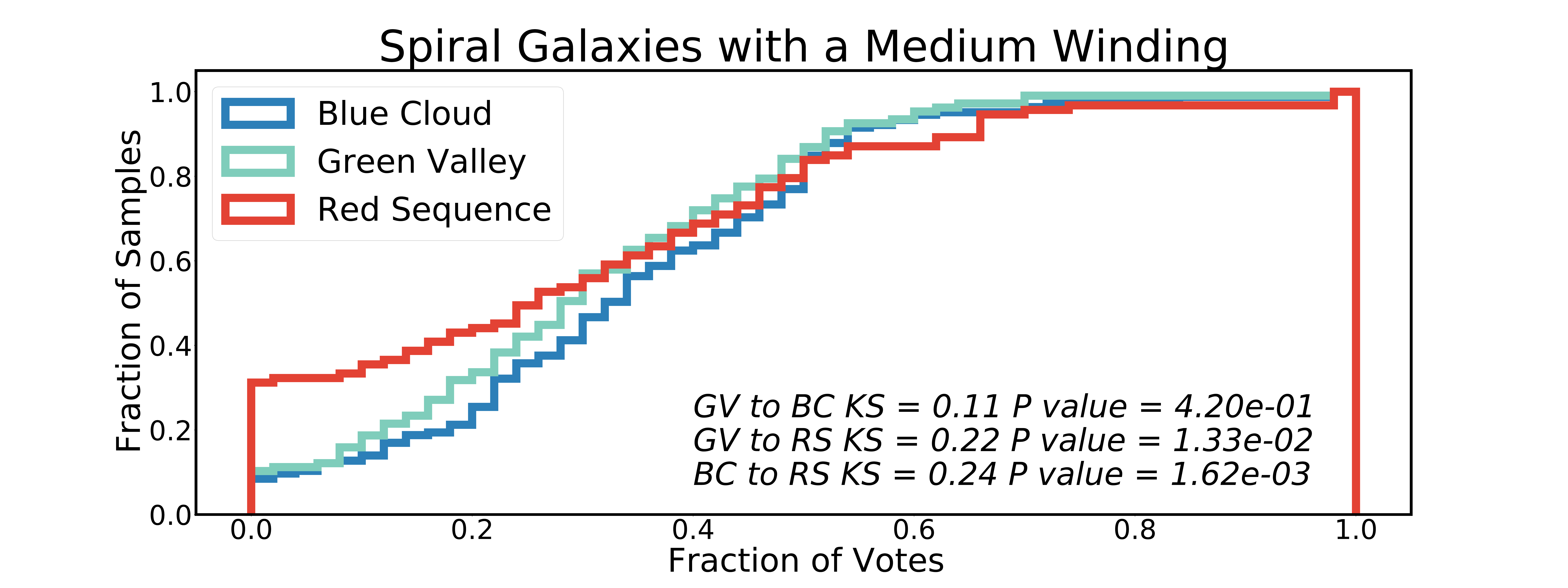}
    \includegraphics[width=0.5\textwidth]{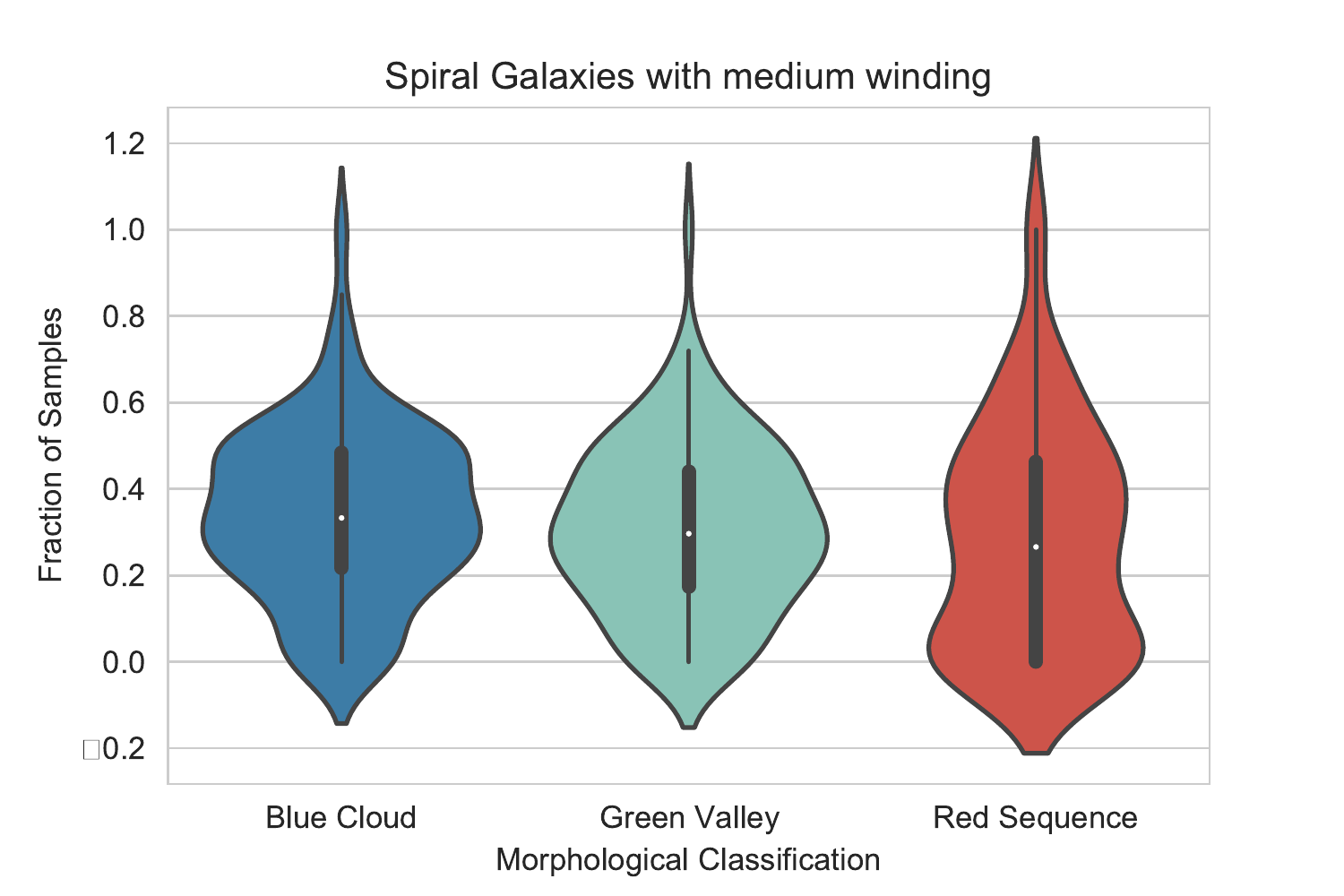}   
    \caption{Histogram of the fraction of votes in favor of the spiral galaxies having a Medium winding  (T05 in the GZ questionnaire). The difference between \GV{} and \BC{} is distinguishable with a KS of 0.11,  but the difference between the \GV{} and \RS{} is  Distinguishable at a KS value of 0.22, and the largest difference is between the behaviour of the \RS{} and the \BC{} with a KS of 0.24. the Significance of these results are \GV{} to \BC{} with p value $4.2\times10^{-01}$, \GV{} to \RS{} with $1.33\times10^{-02}$, and \BC{} to \RS{} with $1.61\times10^{-03}$. This shows again how the behaviours of the \BC{} and \RS{} are opposite with \GV{} maintaining the middle.The violin plots represent the same data as the cumulative histogram in the top panel.}
    \label{fig:hist:medium}
\end{figure}

\begin{figure}
    \centering
    \includegraphics[width=0.5\textwidth]{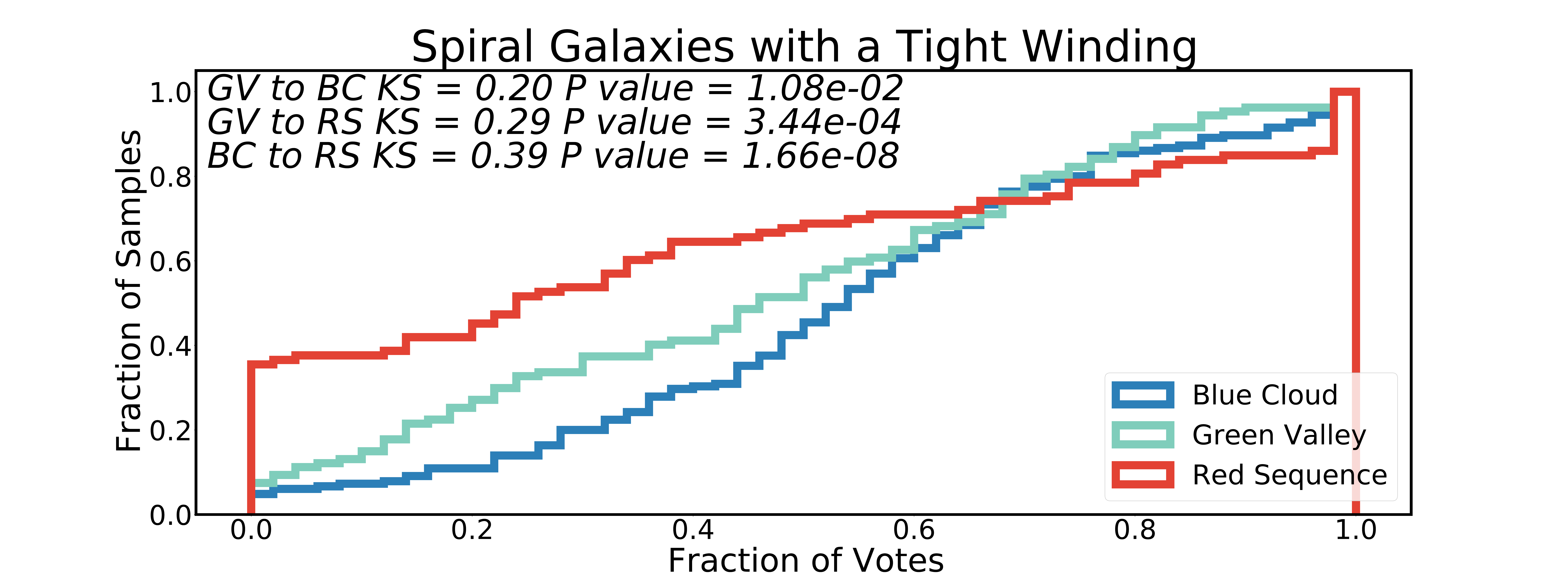}
    \includegraphics[width=0.5\textwidth]{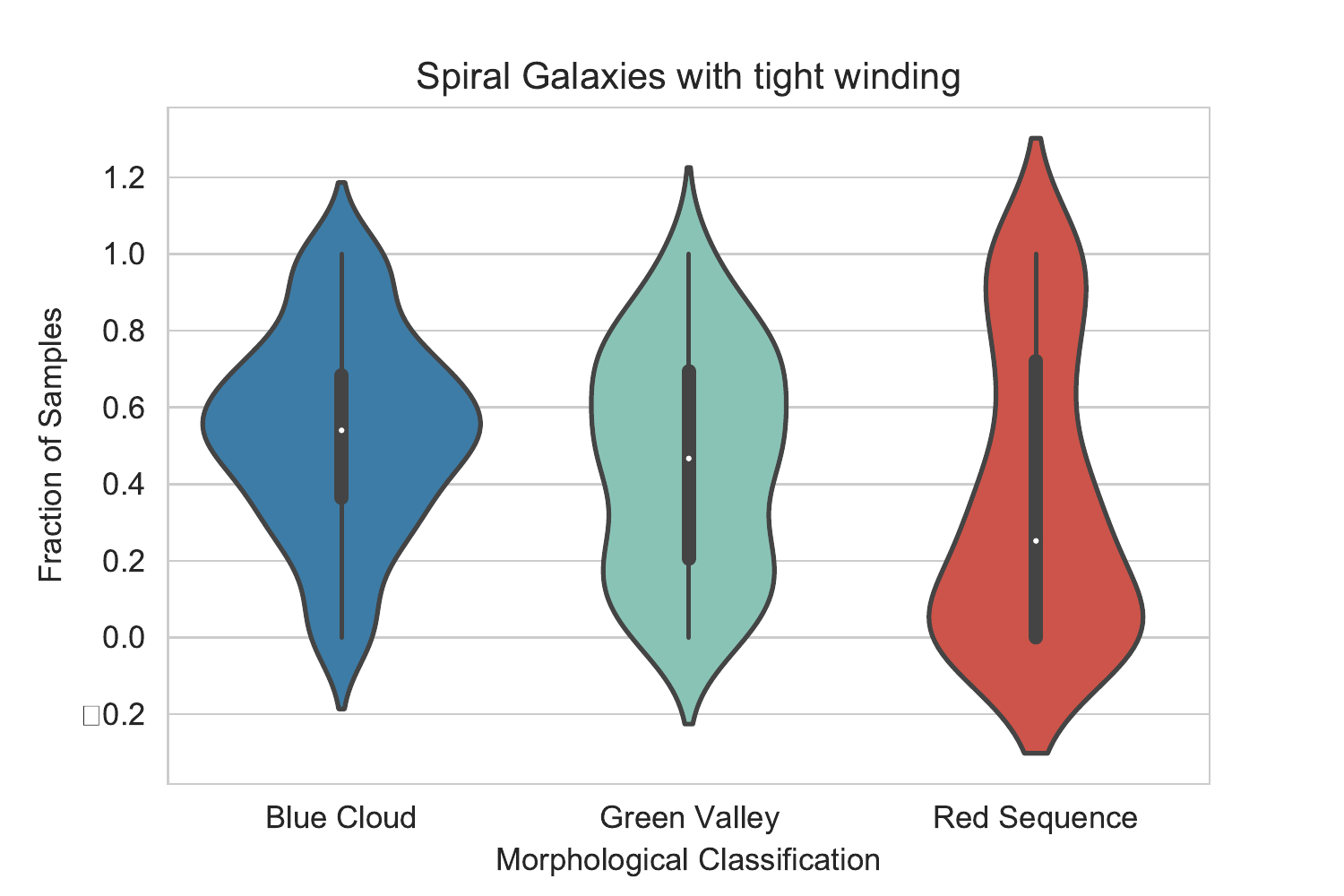}
 \caption{Histogram of the fraction of votes in favor of the spiral galaxies having a tight winding  (T05 in the GZ questionnaire). The difference between \GV{} and \BC{} is distinguishable with a KS of 0.20,  but the difference between the \GV{} and \RS{}.is 0.29 with the difference between the \RS{} and \BC{} being the most distinguishable at 0.39. The significance of these are represented with the p-values of \GV{} and \BC{} at $1.08\times10^{-02}$, \GV{} and \RS{} at $3.44\times10^{-04}$, and \BC{} and \RS{} at $1.66\times10^{-08}$. The violin plots represent the same data as the top panel histograms.}
 \label{fig:hist:tight}
\end{figure}

\subsection{Winding of Spiral Arms}

Figure \ref{fig:gz:questions}, T05: \textit{``How tightly wound do the spiral arms appear?''}

Green valley galaxies tend to follow the \BC{} in behaviour in spiral arm winding of tight and medium, but lead in voting of loose winding. \RS{} continues to show the opposite behaviour of the \BC{} (see, for example, Figures \ref{fig:hist:loose},\ref{fig:hist:medium},and \ref{fig:hist:tight}). 

{Since T05 is a choice between these three questions and one cannot progress without clicking one option, the plots in Figures \ref{fig:hist:loose},\ref{fig:hist:medium},and \ref{fig:hist:tight} are complementary. It shows that loose winding is preferred for \BC{} and tight winding for the \RS{} and the \GV{} voting behaviour is somewhere in between. It also shows that the ``medium'' option voting is much more similar for all three populations as a compromise option but remains less of a preference for the \RS{} galaxies, strongly suggesting that \RS{} galaxies have tightly wound arms. }

\begin{figure}
    \centering
    \includegraphics[width=0.5\textwidth]{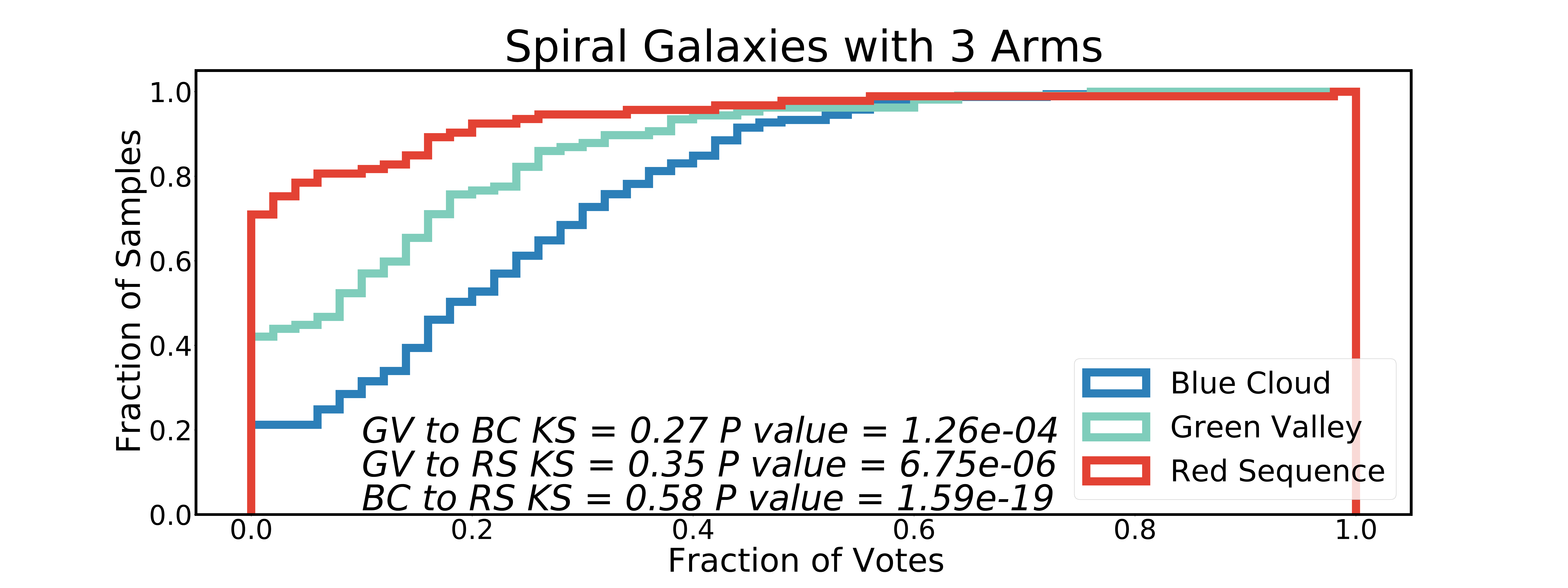}
    \includegraphics[width=0.5\textwidth]{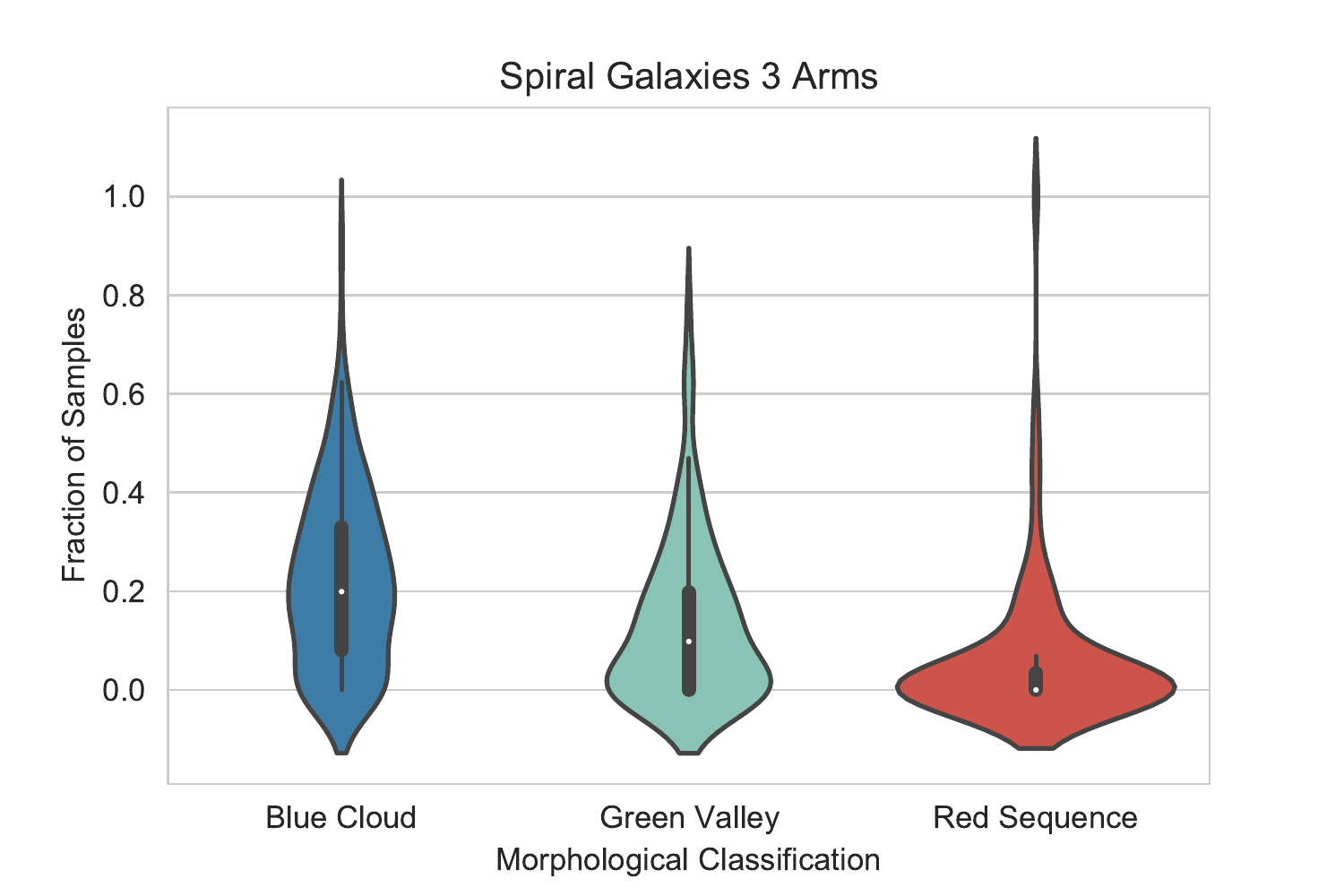}
    \caption{A histogram of the fraction of votes in favor of classifying the galaxies as spiral with 3 arms (T10 in the GZ questionnaire). The difference between the \GV{} and \BC{} is very distinguishable with a K-S value of 0.27. \GV{} to \RS{} is distinguishable with a K-S value of 0.35. The highest difference is seen between \BC{} and \RS{} with a K-S of 0.58. The significance from p-values is $1.26\times10^{-04}$, $6.75\times10^{-06}$, and $1.59\times10^{-19}$ respectively. This shows that spiral galaxies with 3 arms are found more frequent in the \BC{} and \GV{}. The violin plots represent the same data}
    \label{fig:hist:3arm}
\end{figure}

\subsection{Number of Spiral Arms}

Figure \ref{fig:gz:questions}, T06: \textit{``How many spiral arms are there?''}

{Green valley galaxies are more symmetric ($180^\circ$ rotationally symmetric), as the higher relative voting fraction points out. They are favored to have 2 arms, rather than 1 or 3 (which may be $120^\circ$ symmetric). Spiral galaxies with an odd number of arms are more commonly found in the \BC{}, the voting suggests (see, for example, Figure \ref{fig:hist:3arm}). The number of arms have been linked to specific star-formation decline (Porter-Temple et al. submitted) and the relative distribution of voting for 3-armed spirals in the \RS{}, \GV{} and \BC{} reflect this. }

\begin{figure}
    \centering
    \includegraphics[width=0.5\textwidth]{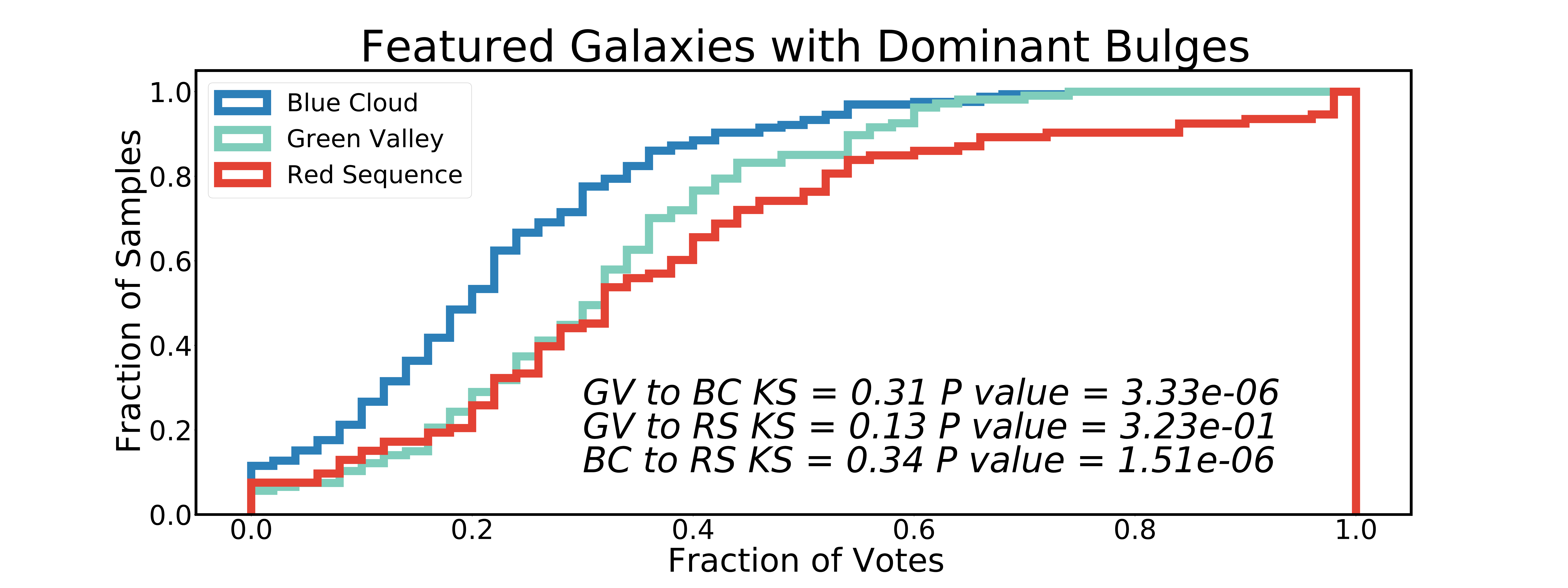}
    \includegraphics[width=0.5\textwidth]{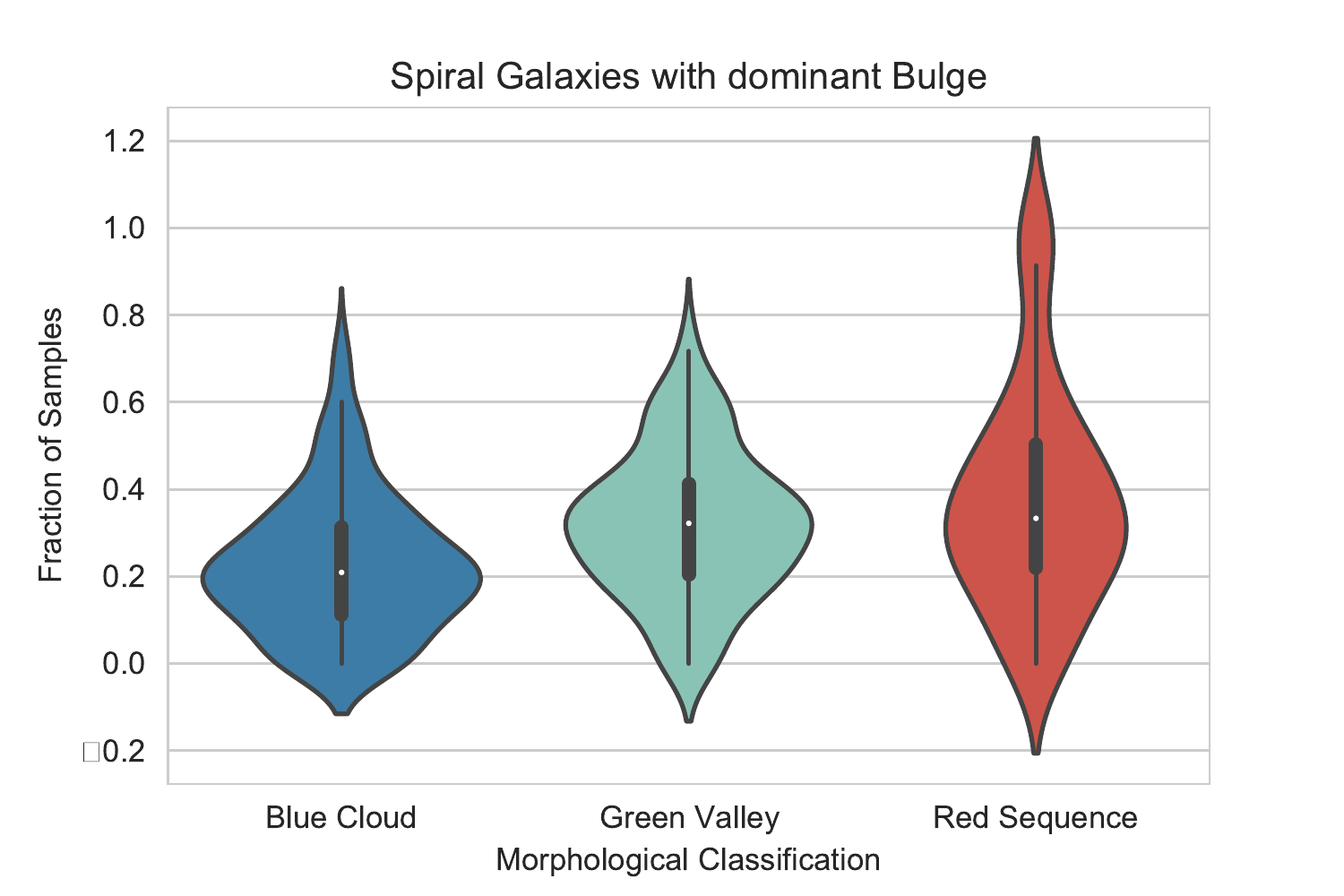}
    \caption{A histogram of the fraction of votes classifying disc galaxies with dominant bulges (T04 in the GZ questionnaire). The difference between the \GV{} and \BC{} is distinguishable with a K-S of 0.31. This difference is less distinguishable between the \GV{} and the \RS{}, with a K-S value of 0.13. When comparing the \BC{} to the \RS{}, the K-S value is 0.34. The significance p-values are $3.33\times10^{-06}$, $3.23\times10^{-01}$, and $7.55\times10^{-03}$ respectively.}
    \label{fig:hist:bulge-dominance}
\end{figure}

\subsection{Central Bulge Prominence}

Figure \ref{fig:gz:questions}, T04: \textit{``How prominent is the central bulge, compared with the rest of the galaxy?"}


{Figure \ref{fig:hist:bulge-dominance} shows the voting behaviour for this question with the highest voting fractions, i.e. largest fraction of the sample with high voting confidence, for a dominant bulge in the \RS{}, followed by \GV{} and \BC{}. This trend is an expected result following the current understanding of galaxy evolution from \GV{} to \RS{} i.e discs fading and turning red and the bulge gaining relative prominence with respect to the disc.
}
\begin{figure}
    \centering
    \includegraphics[width=0.5\textwidth]{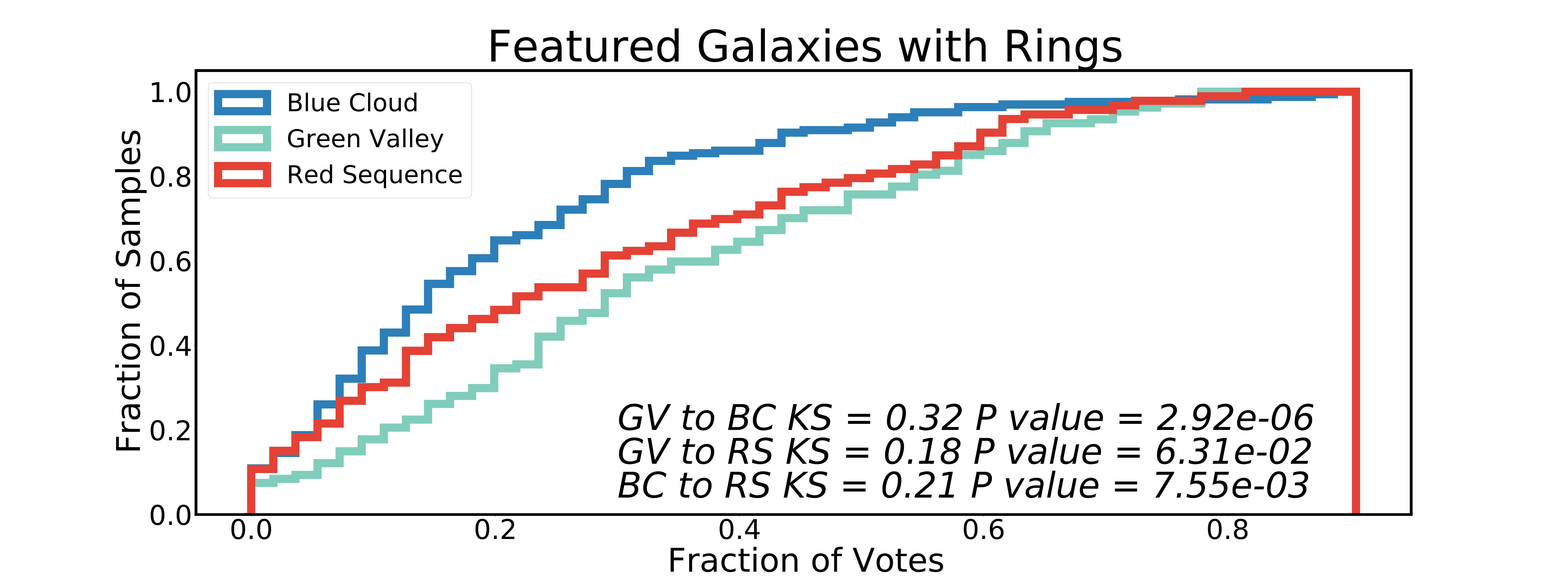}
    \includegraphics[width=0.5\textwidth]{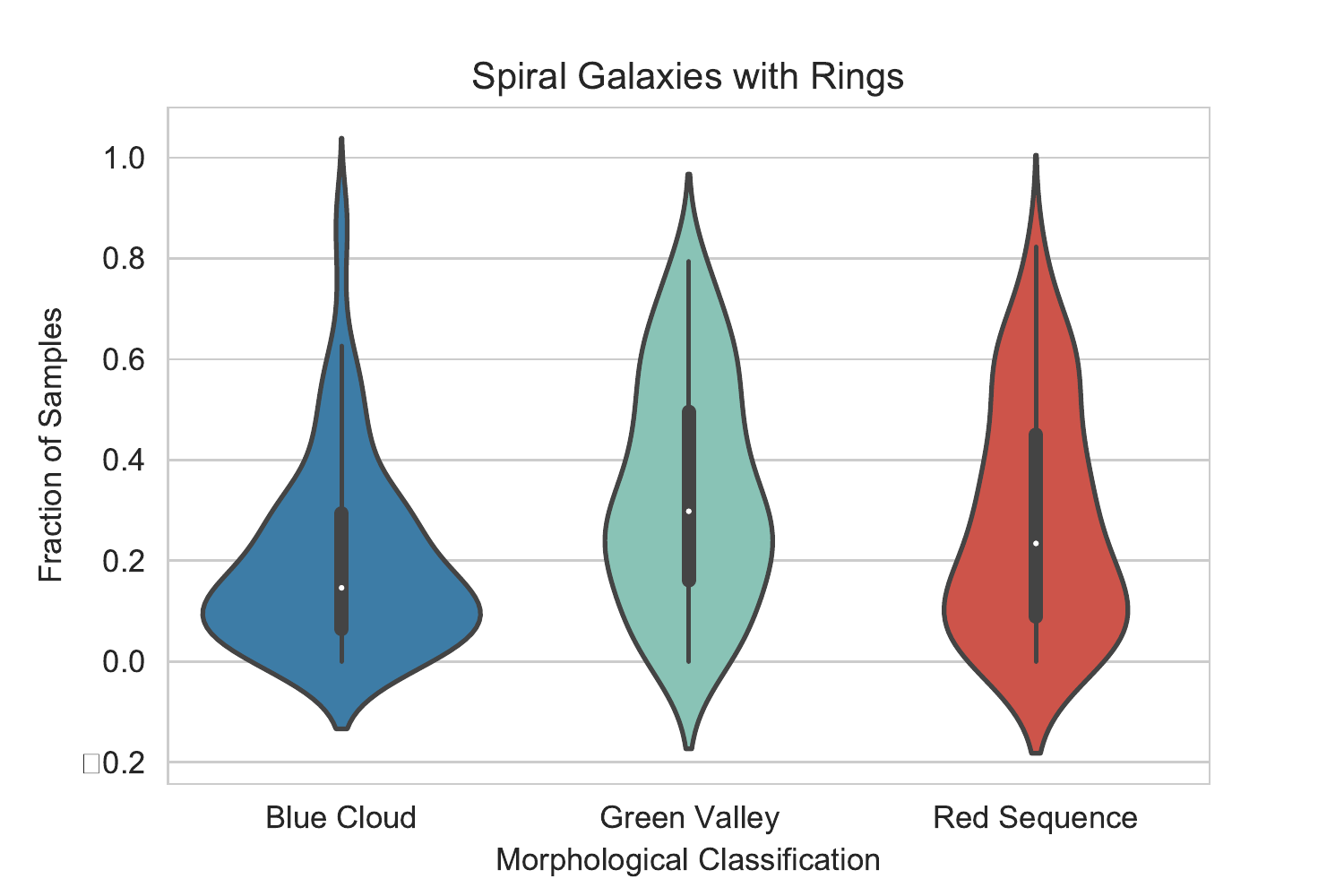}
    \caption{A histogram of the fraction of votes labeling galaxies as featured with rings (T06 in the GZ questionnaire). The \GV{} is distinguishable from the \BC{} with a K-S value of 0.32. The \GV{} continues to differ from the \RS{} with a K-S value of 0.18; while the \RS{} to \BC{} difference has a K-S value of 0.21. The significance p-values are $2.92\times10^{-06}$, $6.31\times10^{-02}$, and $7.55\times10^{-03}$ respectively. This is the first time we observe a difference in the location of the \GV{} data. Here, it is no longer in between the \BC{} and the \RS{}, thus presenting behaviour of its own. }
    \label{fig:hist:rings}
\end{figure}

\subsection{Rings}

\textit{Figure \ref{fig:gz:questions}, T10: ``Do you see any of these odd features in the image?"}

{This last question in the question tree of the Galaxy Zoo (Figure \ref{fig:gz:questions}) is for citizen scientist to identify rarer morphological phenomena, which are expected to be rare, such as accidental overlaps of galaxies along the line-of-sight \citep[see][for a discussion on these]{Keel13,Holwerda17}. One of the options is to identify a ring, as opposed to a lensing object, as this odd feature.} 

Though the source of rings is still heavily debated, the abundance of inside quenching may be a potential cause \citep{Kelvin18}. We expect a direct correlation between the presence of bars and rings in featured galaxies if the inside-out quenching theory holds true. We see this behaviour in the \BC{} and the \RS{} (steady increase in bars in disc galaxies from \BC{} to \RS{}, Figure \ref{fig:hist:bar}). However, there is an even stronger prevalence of ring galaxies in the \GV{} (Figure \ref{fig:hist:rings}). This could be a sign of increased inside quenching or ring formation in the \GV{}, driven by other possible factors.

\section{Discussion}
\label{sec:discussion}

{The transition of galaxies from the star-forming blue cloud to the passive red sequence through the green valley is thought to be from a variety of processes, both internal and external and can be in either direction \citep[see][for a review]{Salim15}. There appears to be a relation between galaxy morphology and their transition speed \citep{Schawinski14,Smethurst15}; smooth galaxies undergo a rapid transit through major mergers, intermediate complex galaxies (e.g. S0) undergo minor mergers and galaxy interactions for an intermediate crossing scale, and disc galaxies cross the slowest due to secular processes.} In disc galaxies, bars and bulges suspected to play a role in the quenching process \citep{Nogueira-Cavalcante17,Gu18,Ge18,Kelvin18}. Green valley quenching appears to be ongoing since $z\sim2$ \citep{Jian20} with a mass-dependence on transition speed and phase \citep{Schawinski14,Angthopo20a}. Higher mass galaxies appear to quench mostly due to lack of gas supply \citep{Das21b} and quenching does seem linked to a lack of Circumgalactic Medium (CGM) \citep{Kacprzak21}. At lower masses, morphological features are thought to influence the quenching speed \citep{Smethurst15}, motivating our morphological characterization of \GV{} galaxies.

Citizen scientists were asked if the galaxies they were looking at were ring galaxies. We have analyzed the \GZ{} votes of galaxies of a mass ($10.25 < log(M_*/M_{\odot}) < 10.75$), as was done in \cite{Kelvin18}. However, we increased the redshift from $z<0.06$ to $z<0.075$ thanks to the improved resolution of the KiDS images. This larger sample size gave us very similar overall results to those from \cite{Kelvin18}, i.e. a higher fraction of ring galaxies in green valley featured galaxies. The data presented in Figure \ref{fig:hist:rings} demonstrates this behaviour. Initially, we had constrained our sample to z $<$ 0.15, the full redshift range of GAMA/GZ data. The results were qualitatively similar but distance effects cannot be fully ruled out and we imposed the $z=0.075$ limit (the distance KiDS resolution corresponds to a 1 kpc feature). 

It is clear that the \GV{} has a higher concentration of ring featured galaxies than both the \RS{} and the \BC{}. This is not the only example of \GV{} exhibiting its own behaviour, as explained in section 3.3, \GV{} have an initial behaviour of looser arm windings (Figure \ref{fig:hist:loose}). 
We have also shown that the \RS{} and \GV{} have higher concentrations of featured galaxies with bars while the \BC{} possesses the lowest amount of featured galaxies with bars (Figure \ref{fig:hist:bar}). 
Earlier studies found that barred galaxies may transition slower \citep{Nogueira-Cavalcante17} or that bulges play a role in the transition through the \GV{} \citep{Ge18}, some of whom may be rejuvenating rather than quenching \citep{Mancini19}.

As previously stated, it is theorized that quenching may cause ring formation. The existence of bars may expedite the quenching process, which may in turn lead to faster ring formation. This shows a possible link between in the role bars and rings in galaxy quenching. The \GV{} represented in the histogram in Figure \ref{fig:hist:bar} shows a difference in behaviour than seen with rings in Figure \ref{fig:hist:rings}. {The voting for bars in \GV{} galaxies in Figure \ref{fig:hist:bar} is in between the voting for \BC{} and \RS{}. The voting in \GV{} in Figure \ref{fig:hist:rings} is more confident in a rings than either the \BC{} or the \RS{}; a larger fraction of the \GV{} sample has a higher confidence in rings than either comparison sample.} 

{However, by looking at our violin graphs portion of Figure \ref{fig:hist:rings}, we see that the distribution of the \GV{} galaxy sample is spread over the whole range of possible values (galaxies with high and low confidence in a ring) while the \BC{} galaxies are mostly clustered at low confidence in rings while the \RS{} galaxies resemble a high confidence and a low confidence population. The \GV{} resemble the \RS{} but with higher voting fractions. }

This could be due to the fact that this is a transition zone and the younger \GV{} galaxies have yet to exhibit the bar or ring behaviours that may occur later in their lifetimes in the \GV{}, just before entering the \RS. Future studies with even more Galaxy Zoo information will help probe the link between bar and ring formation and the green valley population.   


Furthermore, we studied the possibility of a correlation between Dominant bulges and rings (Figures \ref{fig:hist:bulge-dominance} and \ref{fig:hist:rings} respectively). 
Though it does appear that the \GV{} is in the middle in both the bulge and ring distributions, the role of internal quenching remains unclear. 
It is not clear why a fading disc will result in either a ring or a tightening of the spiral arm.
Perhaps, over time, the reduction of gas in the disc results in a lower density for the spiral density wave \citep[e.g.][]{Roberts75, Dobbs14, Shu16b} and thus a different spiral pattern speed. This change in the spiral density wave could lead to either a tightening of the spiral arms or ring formation. This depends however on the dominant formation mechanism for spiral arms \citep[][]{Davis19}. Rings could quench the disc or the quenching of the disc could form rings.



When all conditions are carefully considered, the preference for \GV{} galaxies to be classified with rings suggests that any quenching in the \GV{} is accompanied by subtle changes in disc morphology as well as mere dimming of the disc.

\section{Conclusions} \label{sec:conclusions}



We find that the \RS{} leads in highest concentration of featured galaxies with bulges. It is followed by the \GV{} (Figure \ref{fig:hist:bulge-dominance}). A lack of new-forming stars in the \RS{} leads to a lack of contrast, which may allow bulges to be more visible. This is the diametrically opposite case to bulges in the \BC{}.  

Our results match \GZ{} voting (on KiDS images) in confirming that \GV{} featured galaxies have the most rings in comparison to their \BC{} and \RS{} counterparts (Figure \ref{fig:hist:rings}. This confirms the initial prominence of rings in \GV{} galaxies found by \cite{Kelvin18}. 

Our findings also show a gradual loosening of spiral arms as galaxies enter the \GV{}. \BC{} galaxies are viewed predominantly with tightly wound arms, while every kind of spiral arm winding is present in the \RS{}.
A trend is visible in the voting on central bulge prominence; bulges become more dominant from \BC{} to \RS{}, with the \GV{} displaying an intermediate distribution. 

Our thorough study of galaxies classified in the \GV{} indicates that their behaviours typically share characteristics with both the \RS{} and \BC{}, placing it in the middle (Figure \ref{fig:hist:spiral}), thereby highlighting the transitional interstitial nature of the \GV{}.



\section{Data Availability}

This work is predominantly based on public data from the GAMA survey (available at \url{http://www.gama-survey.org/dr3/}) 
and the \GZ{} catalogue for the GAMA fields, to be released with DR4 (Driver et al. \textit{in prep.})

\section*{Acknowledgements}


The material is supported by NASA Kentucky award No: 80NSSC20M0047 (NASA-REU to L. Haberzettl and D. Smith). D. Smith would like to thank C Nasr for her assistance. 

This publication has been made possible by the participation of all the volunteers in the Galaxy Zoo project. Their contributions are individually acknowledged at \url{http://www.
Galaxy Zoo.org/Volunteers.aspx}. 





\clearpage
\newpage

\appendix
\section{Supplemental Figures}

In this section we supplied the histograms that were represented by the rest of the GAMA questions that were not in the main text. 

    


\begin{figure}
    \centering
    \includegraphics[width=0.5\textwidth]{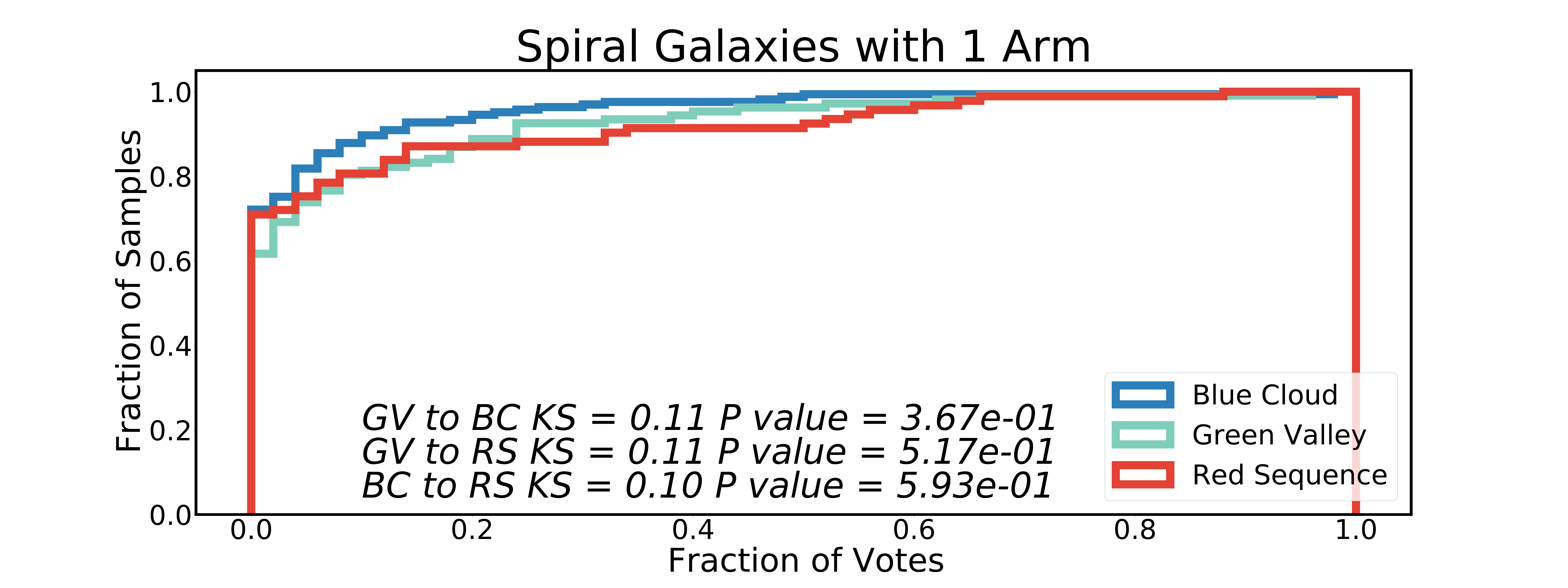}
    \includegraphics[width=0.5\textwidth]{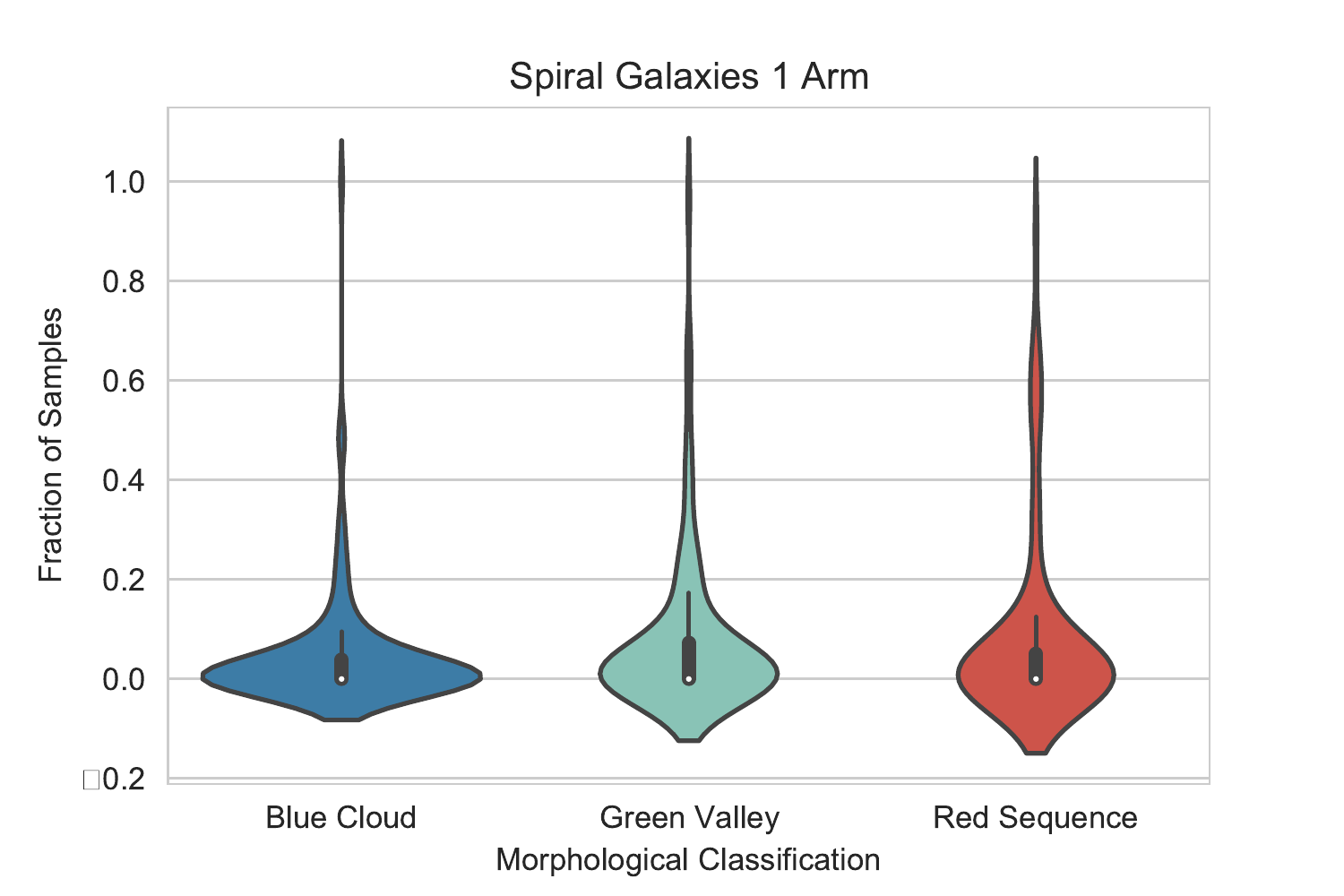}
    \caption{Histogram  of  the  fraction  of  votes  in  favor  of the  galaxies  having  1  Arm  (T10  in  the  \GZ{}  questionnaire). The Difference between the \GV{} and \BC{} is not very distinguishable with a KS of 0.06, \GV{} is more distinguishable from the \RS{} with a KS value of 0.12, keeping \GV{} between the behaviour of the \RS{} and \BC{} that have the most difference with a KS value of 0.18. The significance between the \GV{} and \BC{} are 0.1, \GV{} to \RS{} $4.18\times10^{-05}$, and \BC{} to \RS{} $5.48\times10^{-11}$.}
    \label{fig:hist:1arm}
\end{figure}

\begin{figure}
    \centering
    \includegraphics[width=0.5\textwidth]{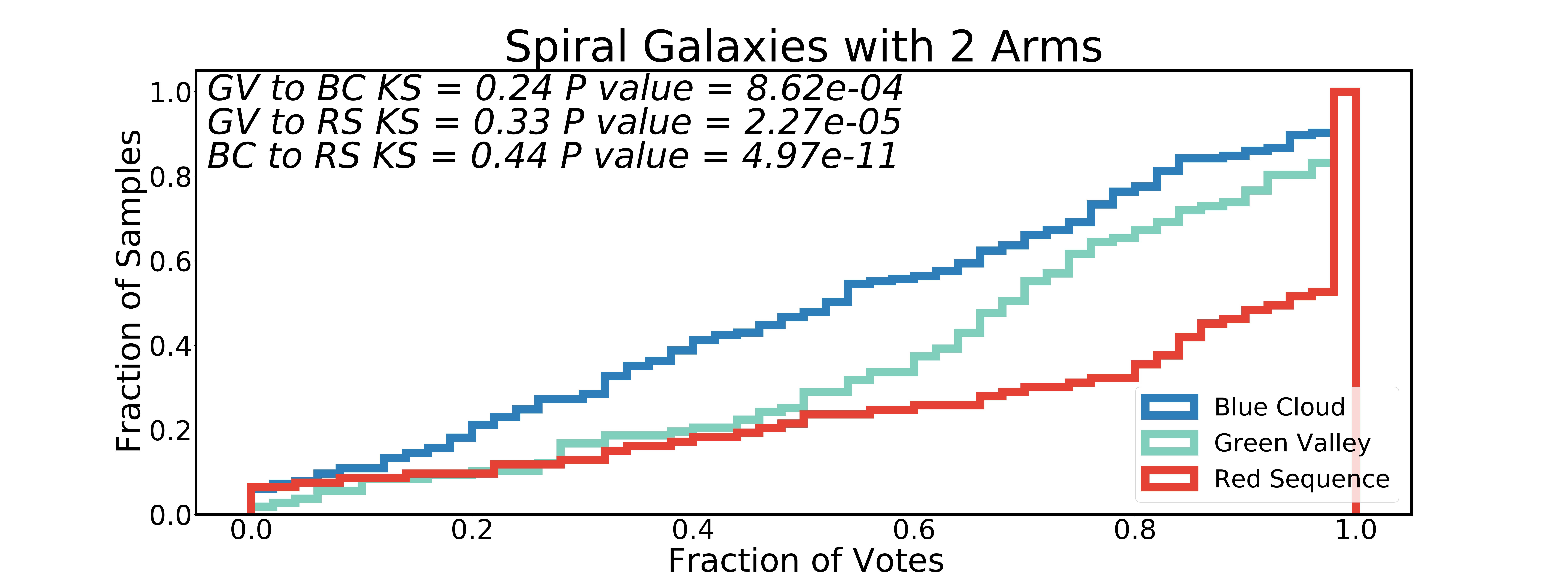}
    \includegraphics[width=0.5\textwidth]{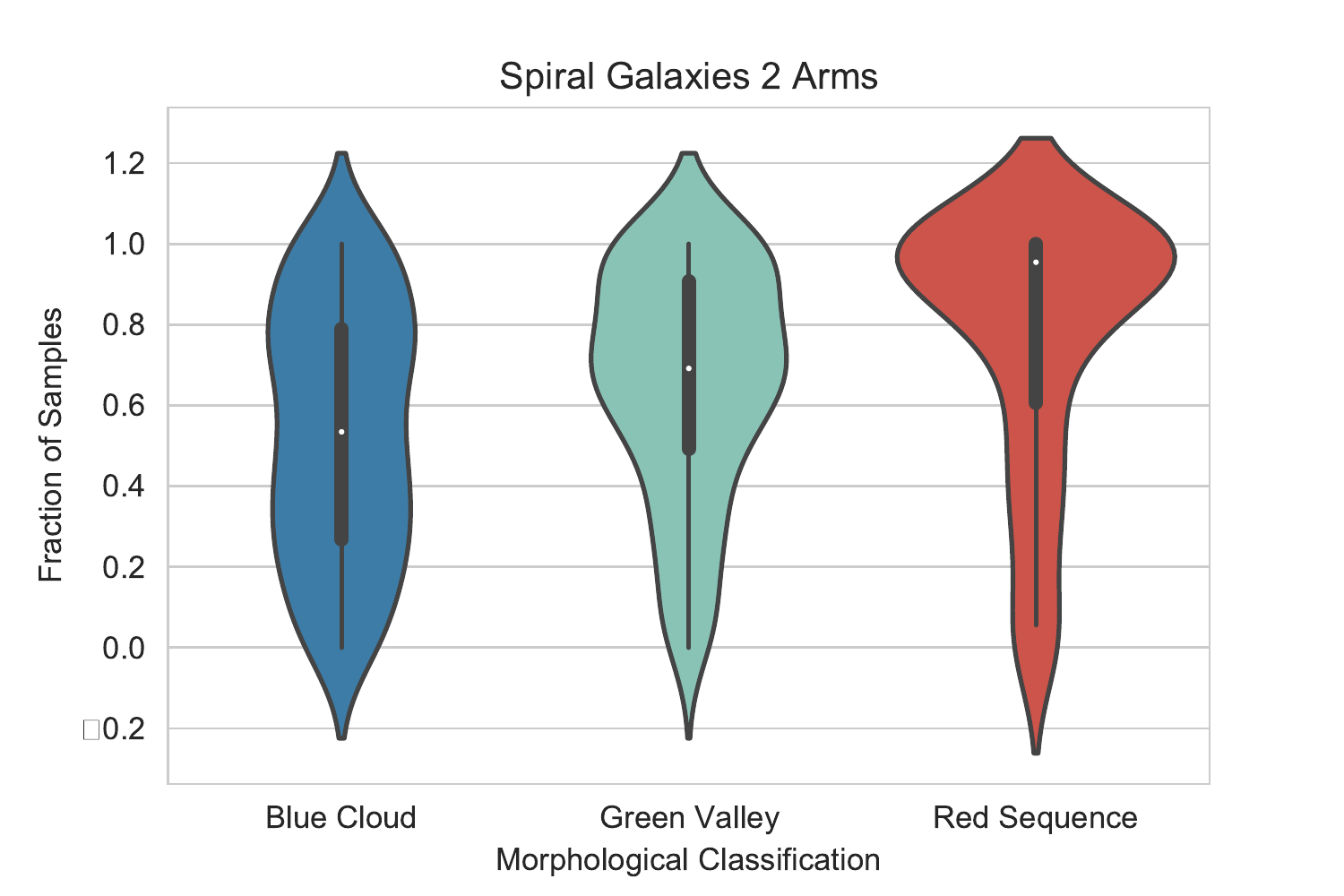}
    \caption{Histogram  of  the  fraction  of  votes  in  favor  of the  galaxies  having  2  Arms  (T10  in  the  \GZ{}  questionnaire). The Difference between the \GV{} and \BC{} is distinguishable with a KS of 0.18, \GV{} is distinguishable form the \RS{} with a KS value of 0.17 staying between the behaviour of the \BC{} and \RS{} that have a KS of 0.33. The significance between the \GV{} and \BC{} are $1.14\times10^{-10}$, \GV{} to \RS{} $1.93\times10^{-09}$, and \BC{} to \RS{} $4.89\times10^{-34}$.}
    \label{fig:hist:2arm}
\end{figure}

\begin{figure}
    \centering
    \includegraphics[width=0.5\textwidth]{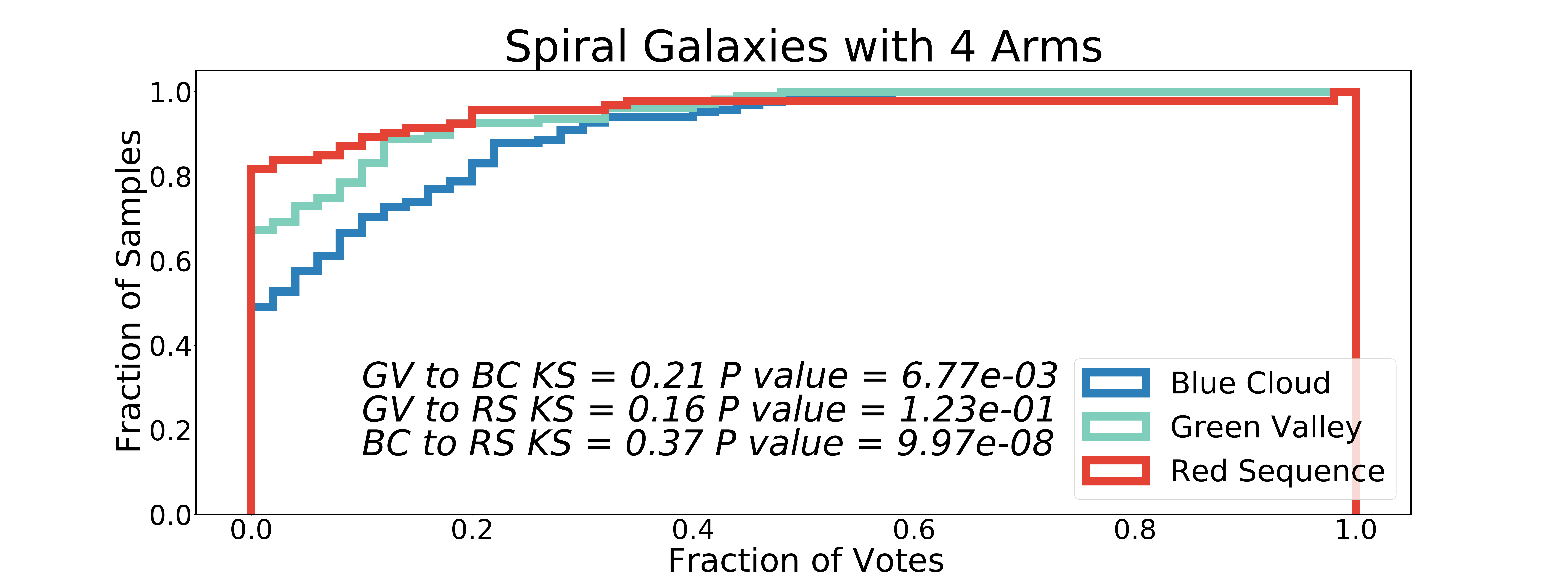}
    \includegraphics[width=0.5\textwidth]{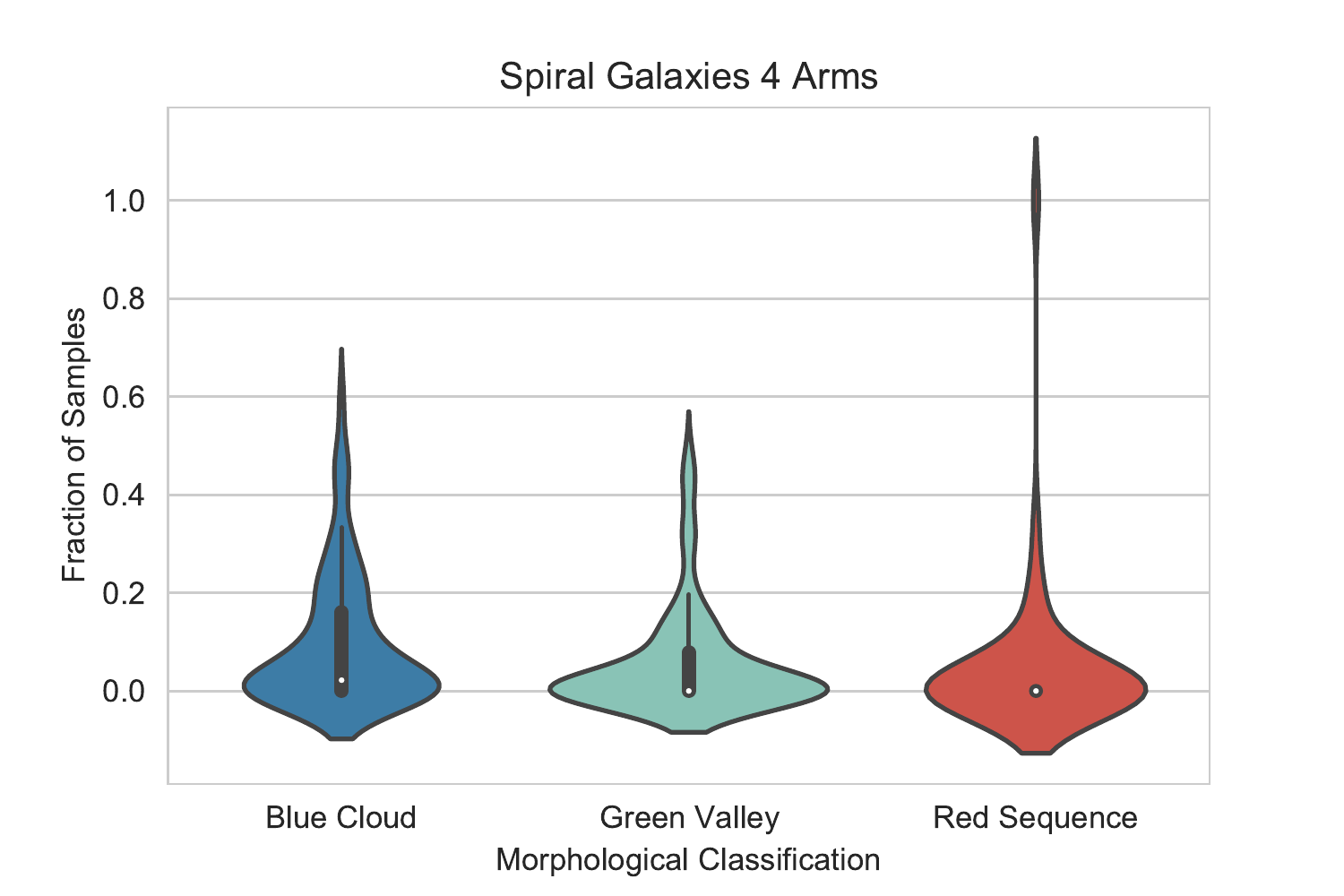}
    \caption{Histogram  of  the  fraction  of  votes  in  favor  of the  galaxies  having  4  Arms  (T10  in  the  \GZ{}  questionnaire). The Difference between the \GV{} and \BC{} is distinguishable with a KS of 0.17, \GV{} is less distinguishable form the \RS{} with a KS value of 0.06 staying between behaviour of the \BC{} and \RS{} with a KS of 0.22. The significance between the \GV{} and \BC{} are $5.32\times10^{-09}$, \GV{} to \RS{} 0.15, and \BC{} to \RS{} $3.5\times10^{-16}$.}
    \label{fig:hist:4arm}
\end{figure}

\begin{figure}
    \centering
    \includegraphics[width=0.5\textwidth]{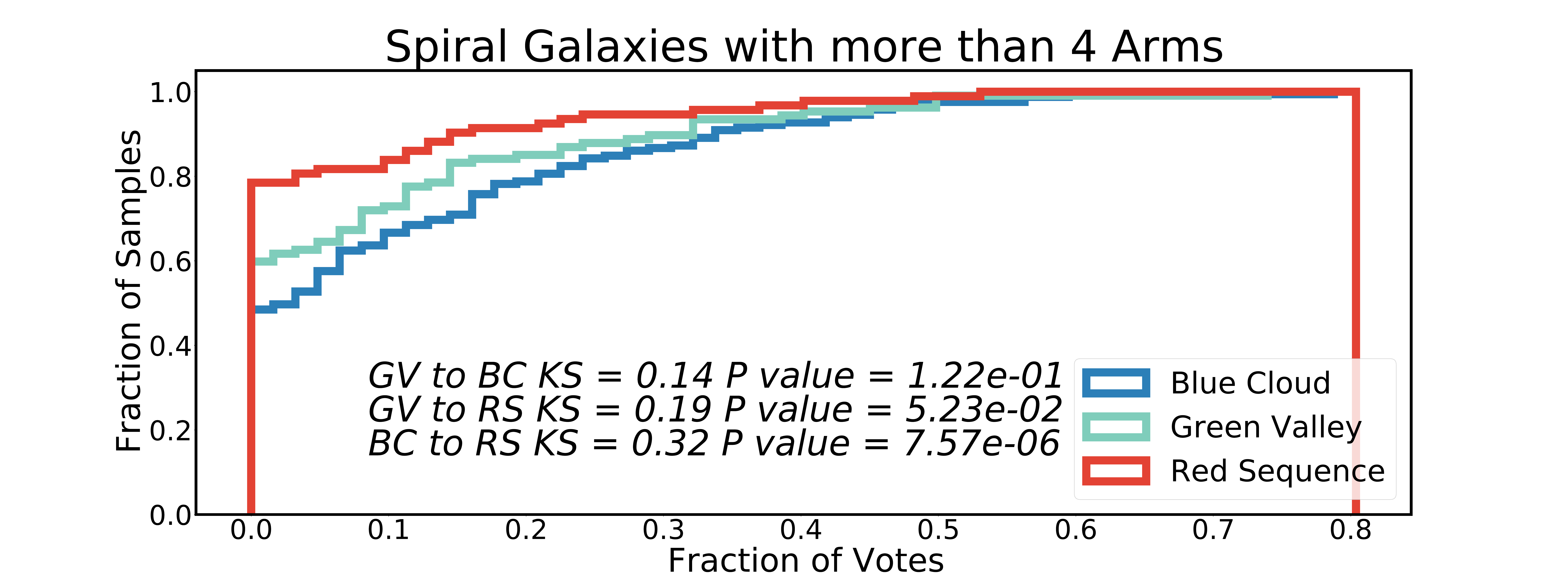}
    \includegraphics[width=0.5\textwidth]{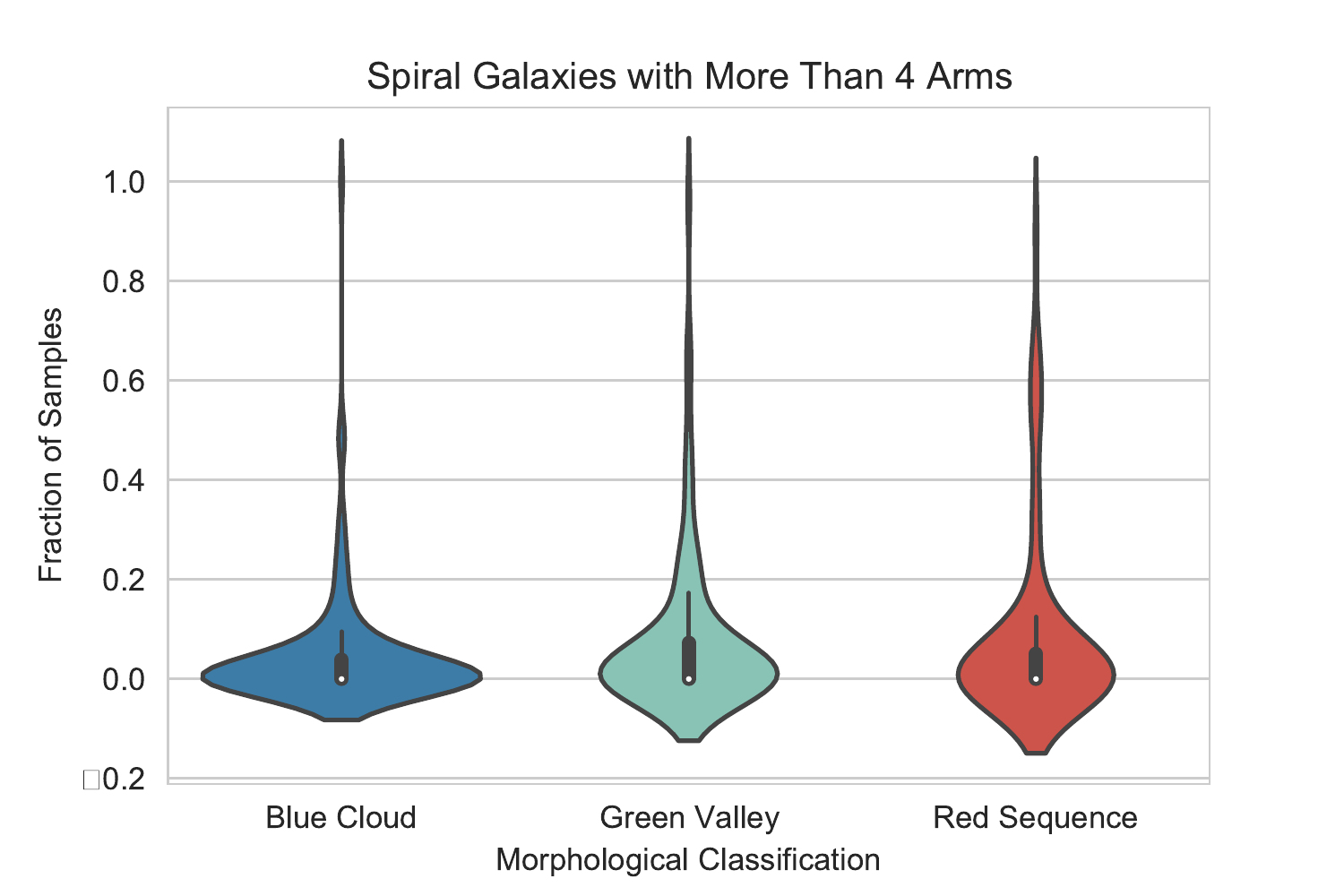}
        \caption{Histogram  of  the  fraction  of  votes  in  favor  of the  galaxies  having more than 4  Arms  (T10  in  the  \GZ{}  questionnaire). The Difference between the \GV{} and \BC{} is distinguishable with a KS of 0.14, \GV{} is less distinguishable form the \RS{} with a KS value of 0.06 , and the KS of \BC{} and \RS{} is 0.20. The significance between the \GV{} and \BC{} are $3.12\times10^{-06}$, \GV{} to \RS{} 0.1, and \BC{} to \RS{} $7.81\times10^{-13}$.}\
        \label{fig:hist:+4arms}
\end{figure}

\begin{figure}
    \centering
    \includegraphics[width=0.5\textwidth]{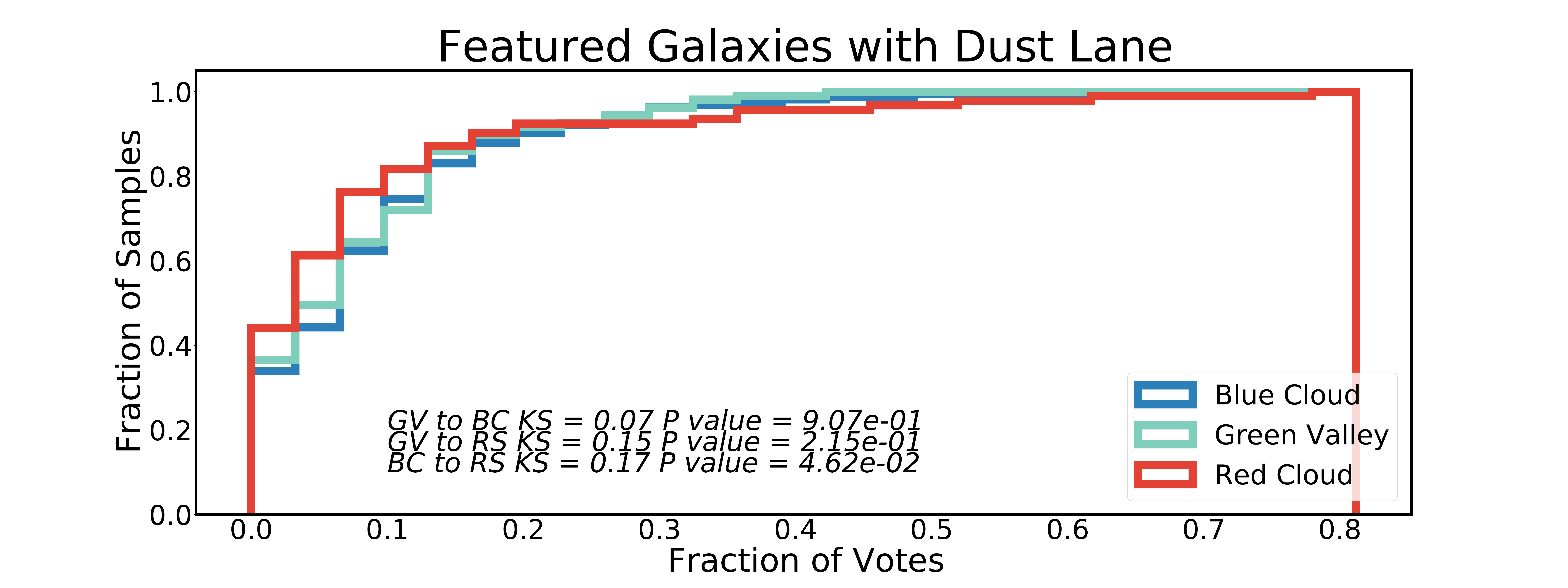}
    \includegraphics[width=0.5\textwidth]{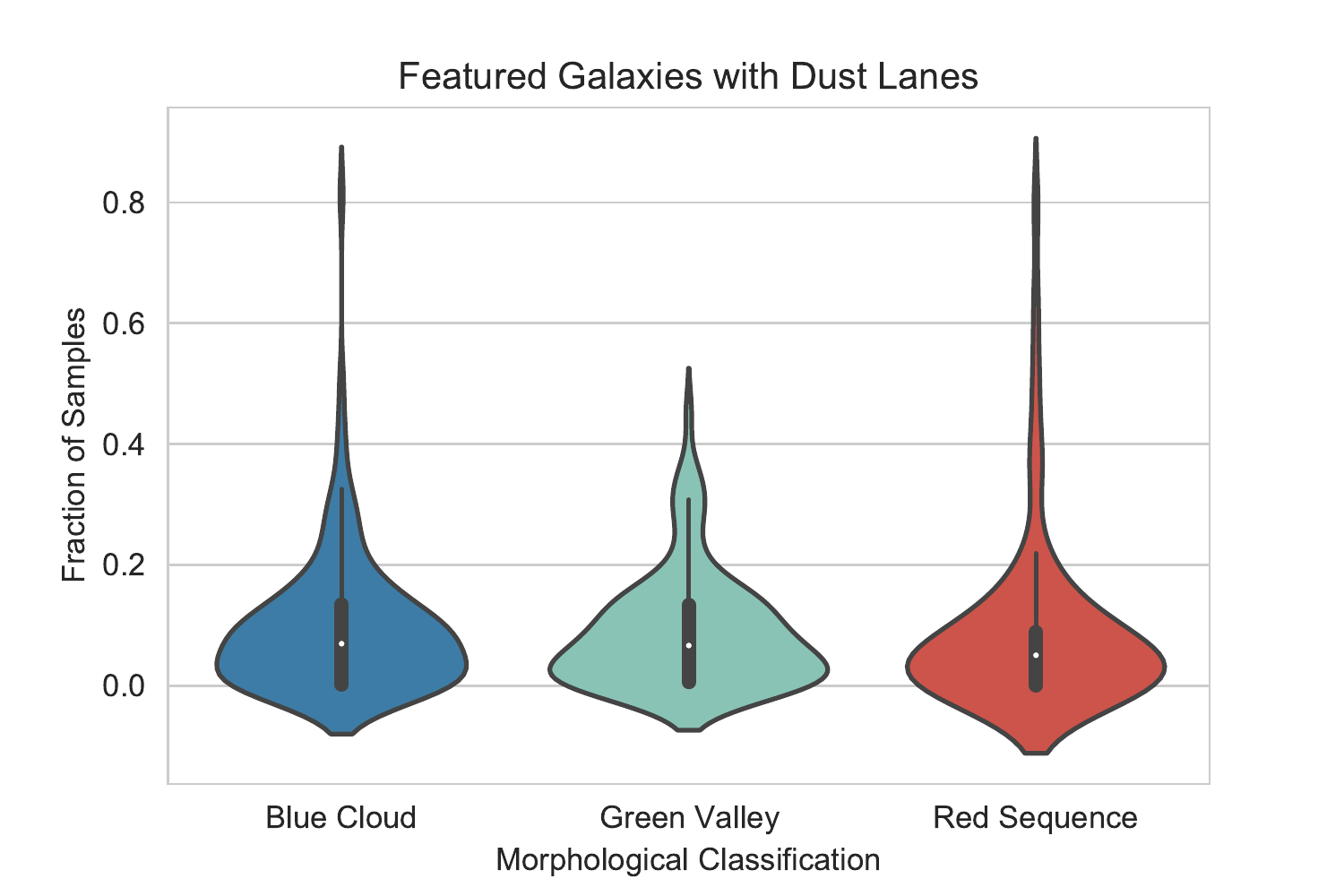}
    \caption{Histogram of the fraction of votes in favor of the galaxies having Dust Lanes (T06 in the \GZ{} questionnaire). The \GV{} distinguishable from the \BC{} and \RS{} with a KS value of 0.08 for both, while \RS{} and \BC{} are distinguishable with a KS value of 0.13. The significance between the \GV{} and \BC{} are 0.02, \GV{} to \RS{} is 0.02, and \BC{} to \RS{} is $5.36\times10^{-06}$.}
   \label{fig:hist:lane}
\end{figure}

\begin{figure}
    \centering
    \includegraphics[width=0.5\textwidth]{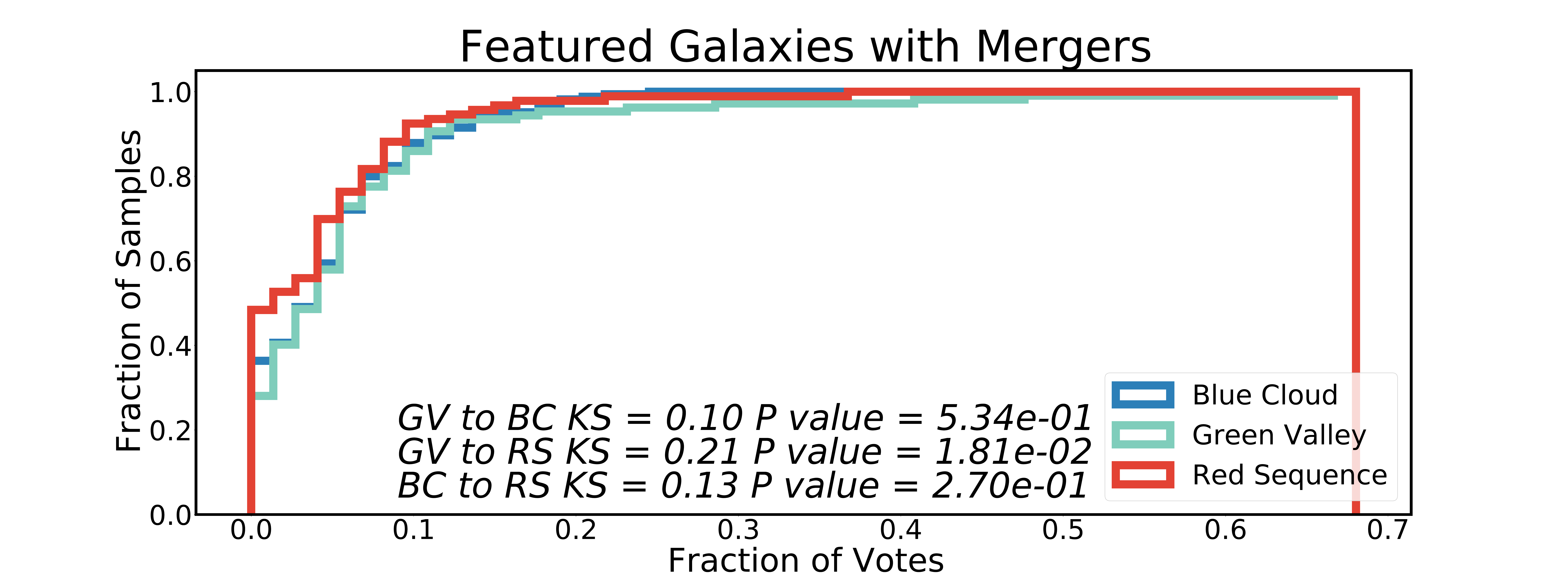}
    \includegraphics[width=0.5\textwidth]{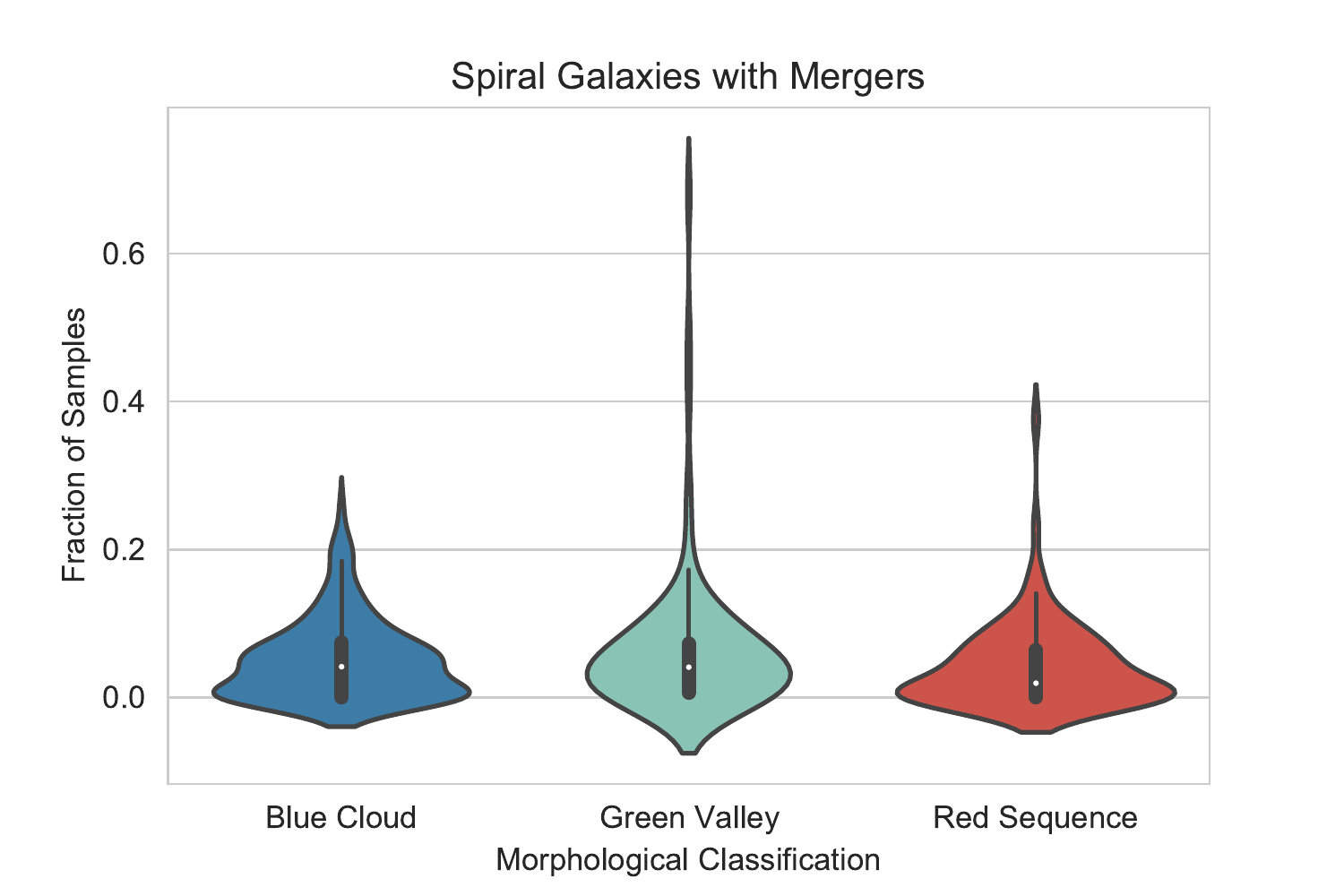}
    \caption{Histogram of the fraction of votes in favor of the galaxies having Mergers (T05 in the \GZ{} questionnaire). The \GV{} is distinguishable from the \BC{} with a KS value of 0.09, \GV{} continues to be differentiate from the \RS{} with a KS value of 0.12 while the \RS{} and \BC{} have a KS Value of 0.18. The significance between the \GV{} and \BC{} are $8.05\times10^{-03}$, \GV{} to \RS{} $6.8\times10^{-05}$, and \BC{} to \RS{} $3.78\times10^{-11}$.}
    \label{fig:hist:merg}
\end{figure}

\begin{figure}
    \centering
    \includegraphics[width=0.5\textwidth]{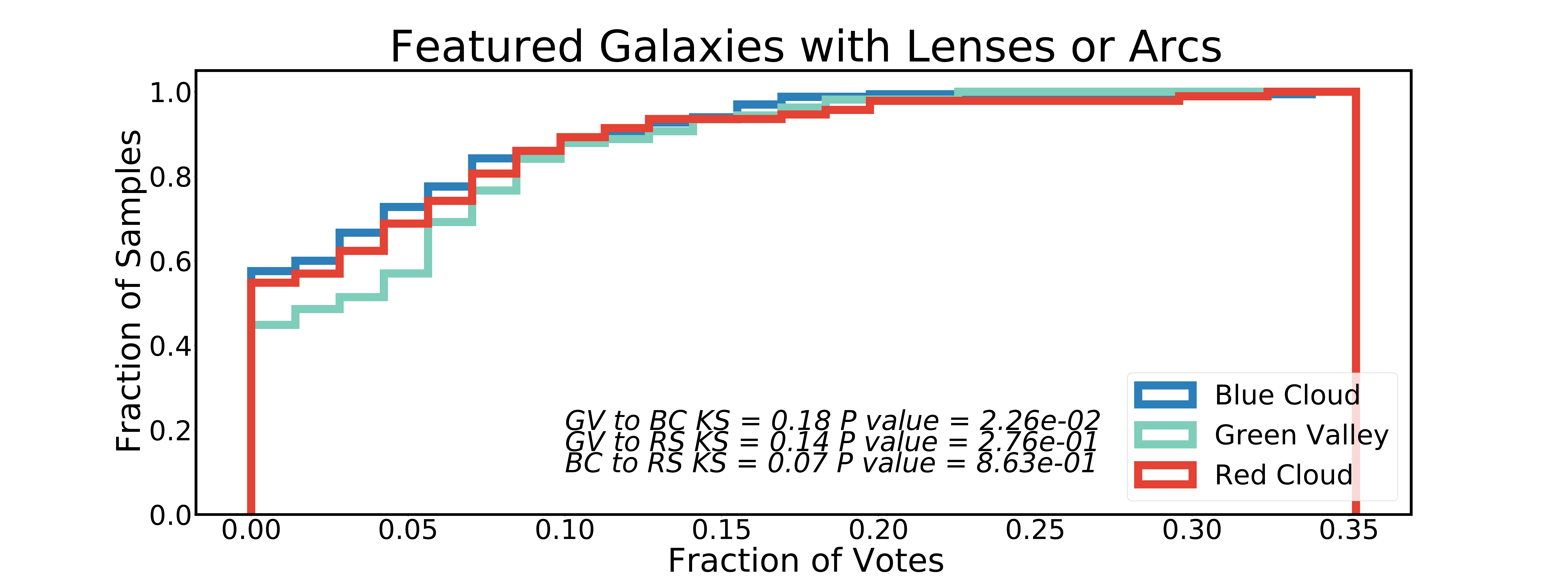}
    \includegraphics[width=0.5\textwidth]{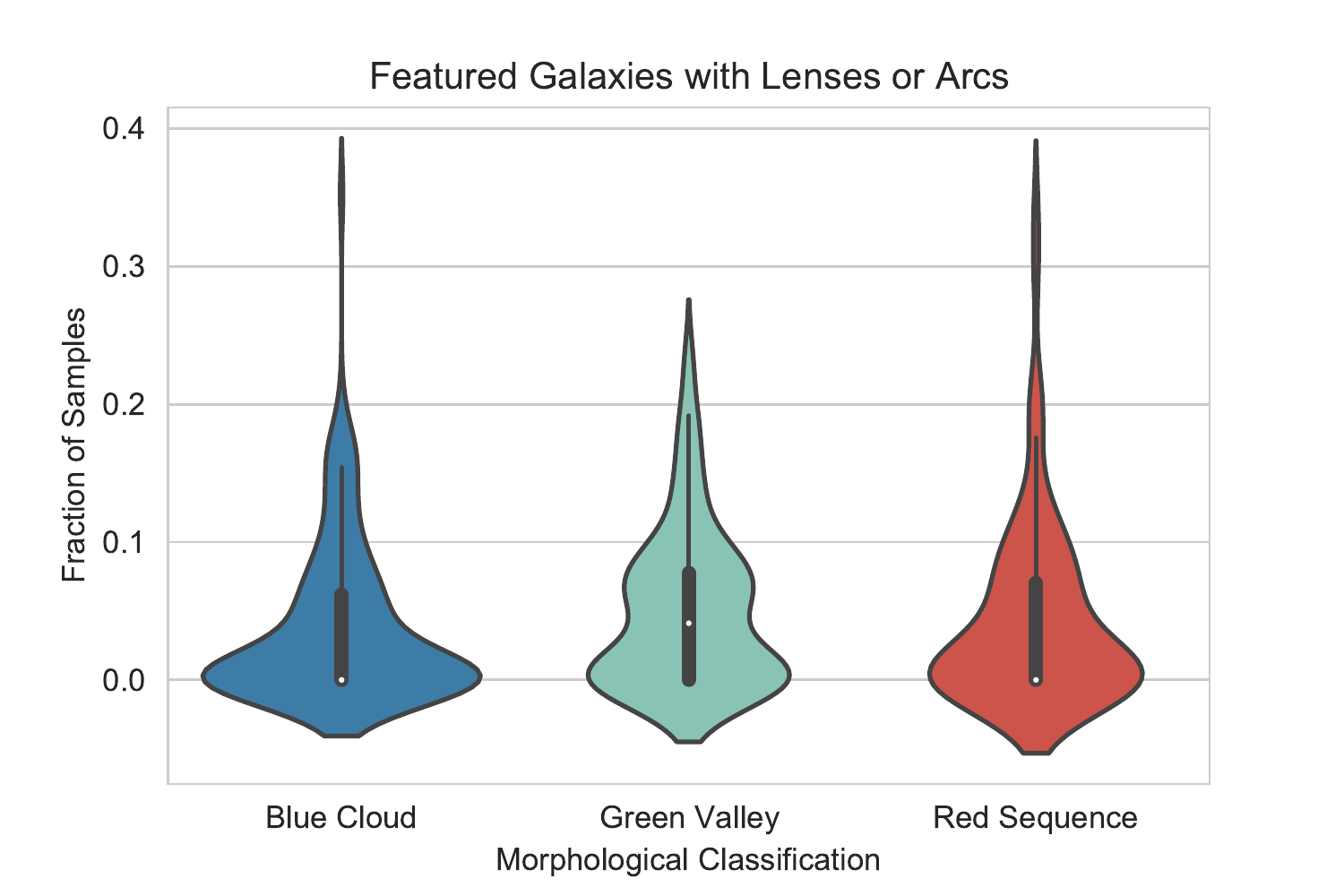}
    \caption{Histogram of the fraction of votes in favor of the galaxies having Lenses or Arcs (T06 in the \GZ{} questionnaire). The \GV{} to \BC{} KS value is 0.04, \GV{} to \RS{} is 0.10, and the \RS{} to \BC{} has a KS Value of 0.09. The significance between the \GV{} and \BC{} are 0.4, \GV{} to \RS{} $1.81\times10^{-03}$, and \BC{} to \RS{} $6.74\times10^{-03}$.}
    \label{fig:hist:lenses}
\end{figure}

\begin{figure}
    \centering
    \includegraphics[width=0.5\textwidth]{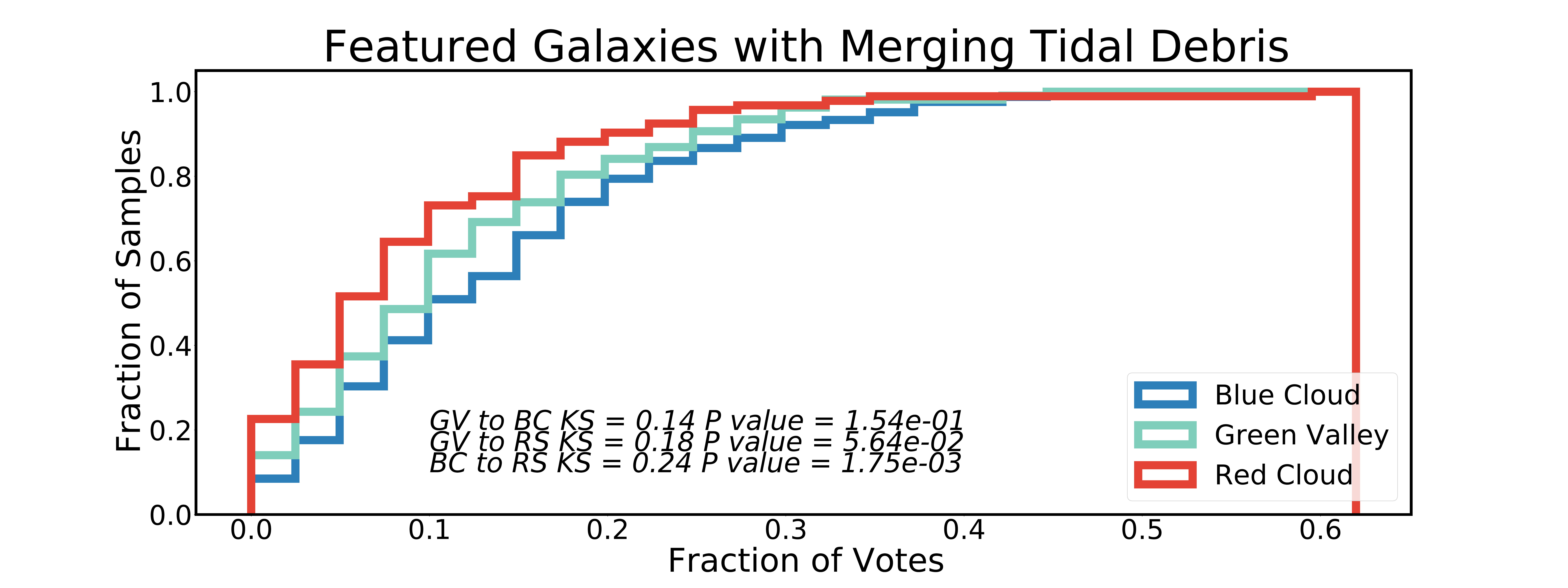}
    \includegraphics[width=0.5\textwidth]{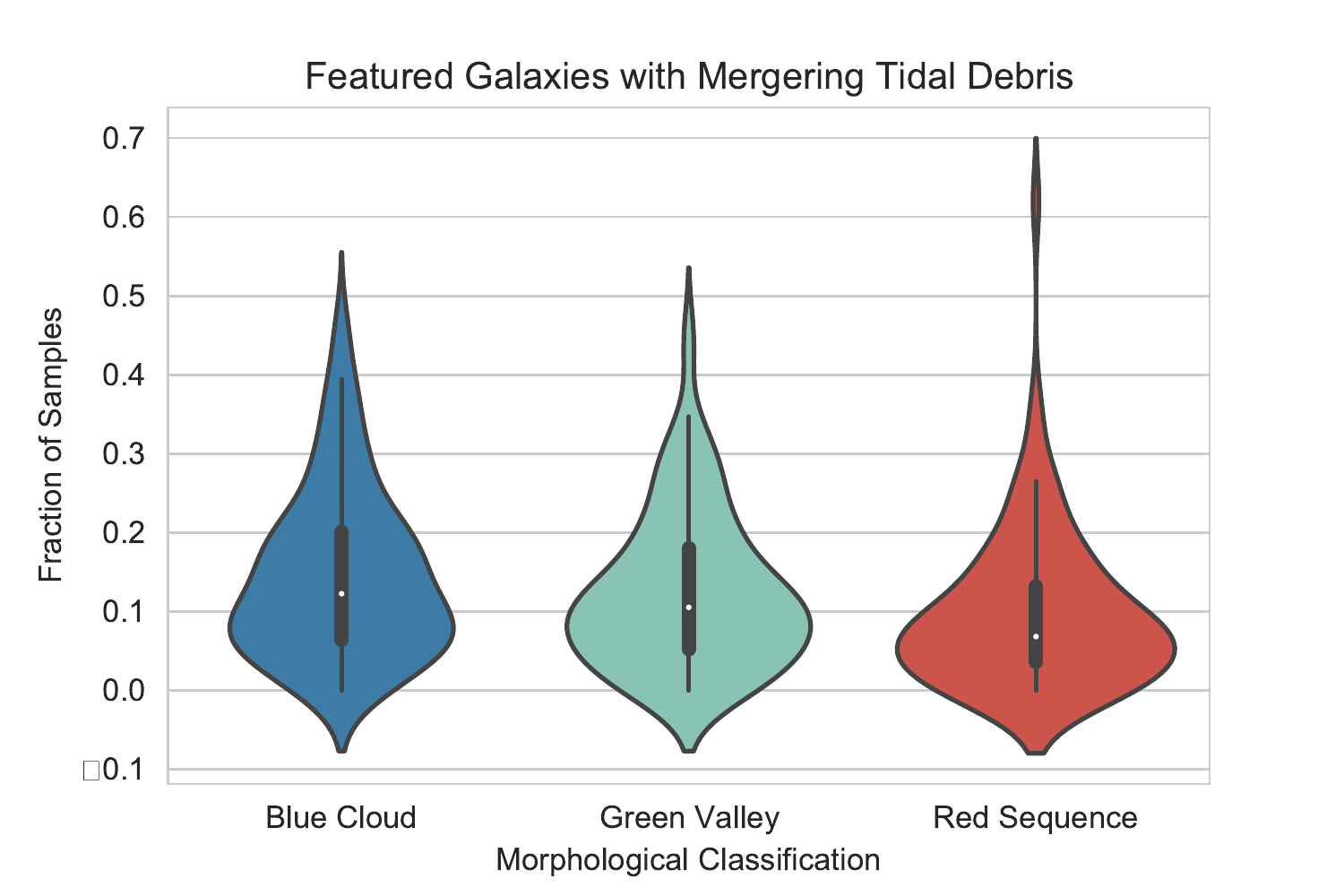}
    \caption{Histogram of the fraction of votes in favor of the galaxies having Tidal Debris (T05 in the \GZ{} questionnaire). The \GV{} is distinguishable from the \BC{} with a KS value of 0.19, \GV{} continues to be differentiate from the \RS{} with a KS value of 0.19 while the \RS{} and \BC{} have a KS Value of 0.35. The significance between the \GV{} and \BC{} are $1.78\times10^{-11}$, \GV{} to \RS{} $5.63\times10^{-12}$, and \BC{} to \RS{} $3.30\times10^{-39}$.}
    \label{fig:hist:debri}
\end{figure}

\begin{figure}
    \centering
    \includegraphics[width=0.5\textwidth]{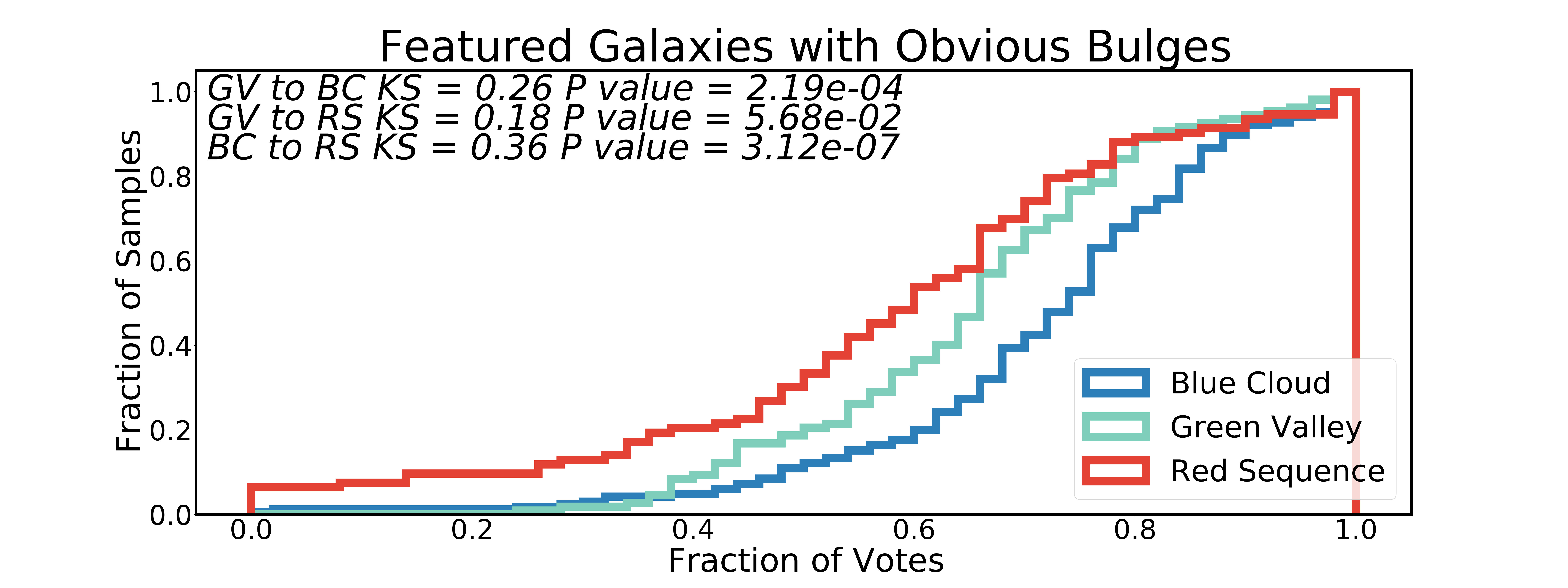}
    \includegraphics[width=0.5\textwidth]{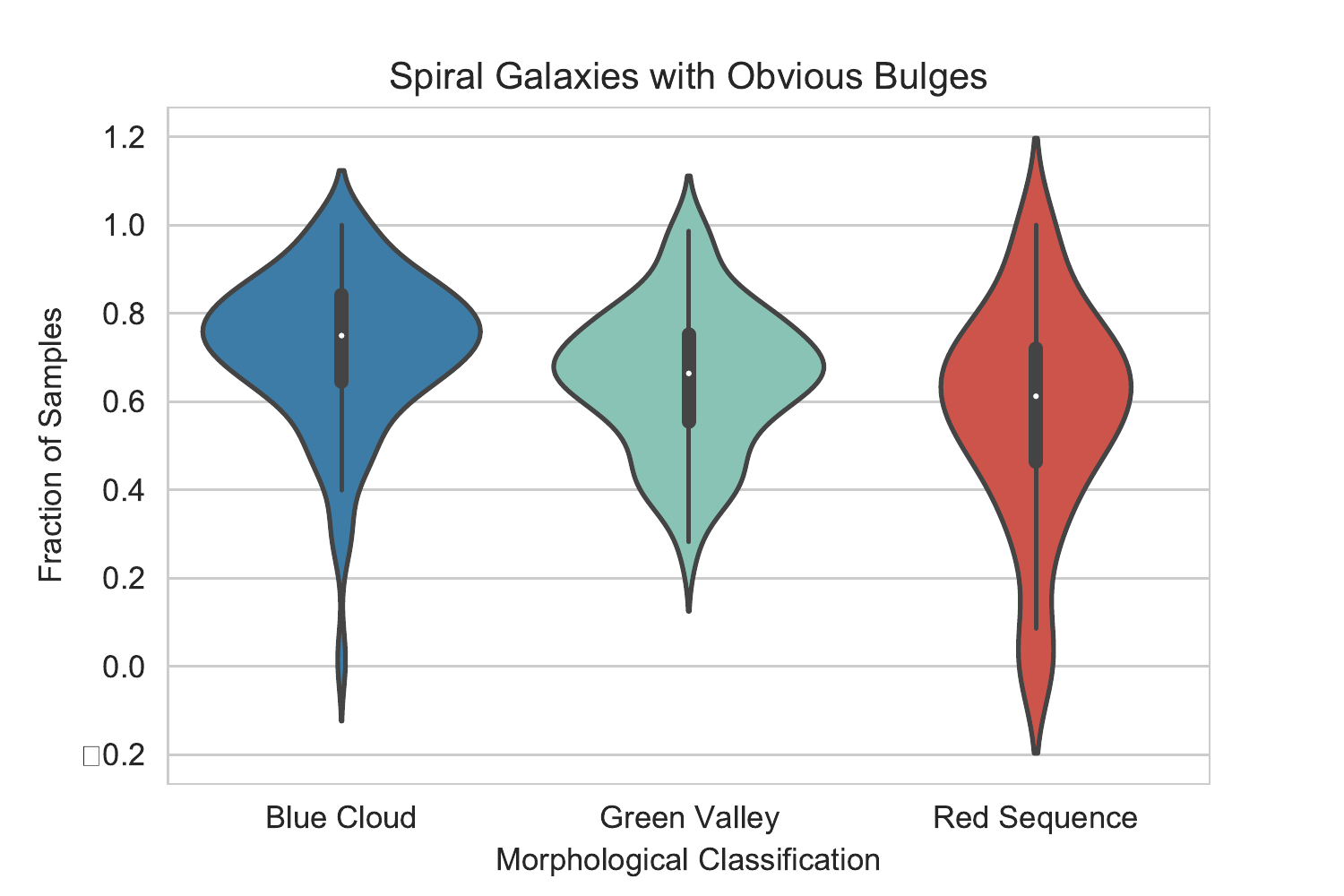}
    \caption{Histogram of the fraction of votes in favor of the galaxies having Obvious Bulges (T04 in the \GZ{} questionnaire). The Difference between the \GV{} and \BC{} is distinguishable with a KS of 0.21,  \GV{} is less distinguishable form the \RS{} with a ks value of 0.11 staying between the behaviour of the \BC{} and \RS{} which have a KS of 0.25. The significance between the \GV{} and \BC{} are $6.64\times10^{-14}$, \GV{} to \RS{} $6.5\times10^{-04}$, and \BC{} to \RS{} $3.66\times10^{-20}$.}
    \label{fig:hist:obvi}
\end{figure}



\begin{table*}
\begin{tabular}{ |p{3cm}||p{3cm}|p{3cm}|p{3cm}|  }
 \hline
 \multicolumn{4}{|c|}{KS Values For Histograms} \\
 \hline
Figure& \GV{} to \BC{} & \GV{} to \RS{} &\BC{} to \RS{}\\
 \hline
\ref{fig:hist:bar}:Bars  & 0.16 &0.10&   0.24 \\
\ref{fig:hist:spiral}:Spiral&  0.32  & 0.20   &0.50\\
\ref{fig:hist:loose}:Loose&0.23 & 0.17&  0.35\\
\ref{fig:hist:medium}:Medium&0.11 & 0.22&  0.24\\
\ref{fig:hist:tight}:Tight& 0.20     & 0.29&0.39\\
\ref{fig:hist:3arm}:3 Arms& 0.27  & 0.35   &0.58\\
\ref{fig:hist:bulge-dominance}:Dominate Bulge& 0.31  & 0.13&0.34\\
\ref{fig:hist:rings}:Rings  & 0.32    &0.18 &   0.21 \\
\ref{fig:hist:1arm}:1 Arm&   0.11  & 0.11   &0.10\\
\ref{fig:hist:2arm}:2 Arm&0.24 & 0.33&  0.44\\
\ref{fig:hist:4arm}:4 Arm&0.21 & 0.16&  0.37\\
\ref{fig:hist:+4arms}:4+ Arms&   0.14  & 0.19 &0.32\\
\ref{fig:hist:lane}:Dust Lane& 0.07  & 0.15   &0.17\\
\ref{fig:hist:merg}:Mergers& 0.10  & 0.21 &0.13\\
\ref{fig:hist:lenses}:Lenses or Arcs  & 0.14    &0.14&   0.07 \\
\ref{fig:hist:debri}:Tidal Debri&   0.18  & 0.14   &0.24\\
\ref{fig:hist:obvi}:Obvious Bulges&0.36 & 0.26&  0.18\\
 \hline
 
\end{tabular}
\end{table*}

\begin{table*}

\begin{tabular}{ |p{3cm}||p{3cm}|p{3cm}|p{3cm}|  }
 \hline
 \multicolumn{4}{|c|}{P Values For Histograms} \\
 \hline
Figure& \GV{} to \BC{} & \GV{} to \RS{} &\BC{} to \RS{}\\
 \hline
\ref{fig:hist:bar}:Bars  &  $6.14\times10^{-02}$&$6.22\times10^{-01}$&   $1.25\times10^{-03}$ \\
\ref{fig:hist:spiral}:Spiral&  $1.39\times10^{-06}$  & $3.12\times10^{-02}$   &$5.66\times10^{-14}$\\
\ref{fig:hist:loose}:Loose&$2.16\times10^{-03}$ & $9.43\times10^{-02}$&  $6.50\times10^{-07}$\\
\ref{fig:hist:medium}:Medium&$4.20\times10^{-01}$ & $1.33\times10^{-02}$&  $1.62\times10^{-03}$\\
\ref{fig:hist:tight}:Tight& $1.08\times10^{-02}$     & $3.44\times10^{-04}$&$1.66\times10^{-08}$\\
\ref{fig:hist:3arm}:3 Arms&$1.26\times10^{-04}$ & $6.75\times10^{-06}$&  $1.59\times10^{-19}$\\
\ref{fig:hist:bulge-dominance}:Dominate Bulge&$3.33\times10^{-06}$& $3.23\times10^{-01}$  &$1.51\times10^{-06}$\\
\ref{fig:hist:rings}:Rings  & $2.92\times10^{-06}$    &$6.31\times10^{-02}$&   $7.55\times10^{-03}$ \\
\ref{fig:hist:1arm}:1 Arm&   $3.67\times10^{-01}$  &   $5.17\times10^{-01}$ &$5.93\times10^{-01}$\\
\ref{fig:hist:2arm}:2 Arm&$8.62\times10^{-04}$ & $2.27\times10^{-05}$&  $4.97\times10^{-11}$\\
\ref{fig:hist:4arm}:4 Arm&$6.77\times10^{-03}$ & $1.23\times10^{-01}$ & $9.97\times10^{-08}$\\
\ref{fig:hist:+4arms}:4+ Arms&$1.22\times10^{-01}$& $5.23\times10^{-02}$& $7.57\times10^{-06}$\\
\ref{fig:hist:lane}:Dust Lane& $9.07\times10^{-01}$  & $2.15\times10^{-01}$   &$4.62\times10^{-02}$\\
\ref{fig:hist:merg}:Mergers& $5.34\times10^{-01}$  & $1.81\times10^{-02}$&$2.70\times10^{-01}$\\
\ref{fig:hist:lenses}:Lenses or Arcs  & $1.54\times10^{-01}$   &$2.76\times10^{-01}$&   $8.63\times10^{-01}$ \\
\ref{fig:hist:debri}:Tidal Debri&   $1.54\times10^{-01}$  & $5.64\times10^{-02}$   &$1.75\times10^{-03}$\\
\ref{fig:hist:obvi}:Obvious Bulges&$2.19\times10^{-04}$& $5.68\times10^{-02}$&$3.12\times10^{-07}$\\

 \hline
\end{tabular}
\end{table*}

\bsp	
\label{lastpage}
\end{document}